\documentclass[sigconf,screen]{acmart}

\setcopyright{cc}
\setcctype{by}
\acmDOI{10.1145/3832783.3837522}
\acmYear{2026}
\copyrightyear{2026}
\acmISBN{979-8-4007-2882-2/2026/10}
\acmConference[ASE '26]{Proceedings of the 41st IEEE/ACM International Conference on Automated Software Engineering}{October 12--16, 2026}{Munich, Germany}
\acmBooktitle{Proceedings of the 41st IEEE/ACM International Conference on Automated Software Engineering (ASE '26), October 12--16, 2026, Munich, Germany}
\acmSubmissionID{ase26main-p2100-p}
\received{2026-03-26}
\received[accepted]{2026-06-18}

\usepackage{afterpage}
\usepackage{url}
\usepackage{ifthen}
\usepackage{listings}
\definecolor{codegreen}{rgb}{0,0.6,0}
\definecolor{codegray}{rgb}{0.5,0.5,0.5}
\definecolor{codepurple}{rgb}{0.58,0,0.82}
\definecolor{backcolour}{rgb}{0.94,0.94,0.94}  
\usepackage{subcaption}
\usepackage{array}
\usepackage{booktabs}
\usepackage{fontawesome5}
\usepackage{multirow}
\usepackage{graphicx} 
\usepackage[table]{xcolor}
\usepackage{subcaption}
\usepackage{colortbl}
\usepackage{makecell}

\usepackage{enumitem}

\usepackage{adjustbox}

\usepackage{subcaption}
\usepackage[most]{tcolorbox}
\newcommand{\rqbox}[1]{

\begin{tcolorbox}[tile, size=fbox, boxsep=1.5mm, boxrule=0pt, top=0pt, bottom=0pt,
borderline west={0.8mm}{0pt}{gray!50!white}, colback=gray!10!white]
\small#1
\end{tcolorbox}
}
\usepackage{caption}

\newcommand{\takeaway}[1]{%
  \noindent\textbf{\textit{Takeaway:}} #1
}

\newboolean{hidediff}
\setboolean{hidediff}{true}
\ifthenelse{\boolean{hidediff}}
{\newcommand{\hl}[1]{#1}}  
{\newcommand{\hl}[1]{\textcolor{blue}{#1}}}  

\usepackage{tikz}
\newcommand{\blackcircle}[1]{%
    \tikz[baseline=(char.base)]{
        \node[shape=circle, fill=black, text=white, inner sep=1pt] (char) {#1};
    }%
}

\begin{document}
\title[Demystifying Solana Bots: From GitHub Blueprints to On-Chain Fingerprints]{Demystifying Solana Bots: From GitHub Blueprints\\to On-Chain Fingerprints}

\author{Xiaoye Zheng}
\orcid{0009-0001-9048-1930}
\affiliation{%
  \institution{Zhejiang University}
  \city{Hangzhou}
  \country{China}
}
\email{xiaoyez@zju.edu.cn}

\author{Yujing Chen}
\orcid{0009-0005-9690-9183}
\affiliation{%
  \institution{Zhejiang University}
  \city{Hangzhou}
  \country{China}
}
\email{chenyujing@zju.edu.cn}

\author{Minghao Wu}
\orcid{0009-0003-7208-3992}
\affiliation{%
  \institution{Zhejiang University}
  \city{Hangzhou}
  \country{China}
}
\email{wuminghao@zju.edu.cn}

\author{David Lo}
\orcid{0000-0002-4367-7201}
\affiliation{%
  \institution{Singapore Management University}
  \city{Singapore}
  \country{Singapore}
}
\email{davidlo@smu.edu.sg}

\author{Difan Xie}
\orcid{0009-0005-5777-1679}
\affiliation{%
  \institution{Hangzhou High-Tech Zone (Binjiang) Institute of Blockchain and Data Security}
  \city{Hangzhou}
  \country{China}
}
\email{xiedifan@bcds.org.cn}

\author{Daoyuan Wu}
\orcid{0000-0002-3752-0718}
\affiliation{%
  \institution{Lingnan University}
  \city{Hong Kong}
  \country{Hong Kong}
}
\email{daoyuanwu@ln.edu.hk}

\author{Xiaohu Yang}
\orcid{0000-0003-4111-4189}
\affiliation{%
  \institution{Zhejiang University}
  \city{Hangzhou}
  \country{China}
}
\email{yangxh@zju.edu.cn}

\author{Zhiyuan Wan}
\authornote{Zhiyuan Wan is the corresponding author; also with the Hangzhou High-Tech Zone (Binjiang) Institute of Blockchain and Data Security.}
\correspondingauthor
\orcid{0000-0001-7657-6653}
\affiliation{%
  \institution{Zhejiang University}
  \city{Hangzhou}
  \country{China}
}
\email{wanzhiyuan@zju.edu.cn}
\titlenote{This research was supported by the National Science Foundation of China (No. 62472383), the Fundamental Research Funds for the Central Universities (No. 226-2025-00004), and the Open Research Fund of the State Key Laboratory of Blockchain and Data Security, Zhejiang University.}

\begin{abstract}
Solana is an emerging blockchain platform designed for high throughput and low transaction fees, making it inexpensive to submit transactions at scale and, consequently, increasing exposure to bot spamming and related financial exploitation.
Solana bots are typically off-chain software systems that operate in a competitive on-chain execution environment by constructing and submitting transactions, and the bot-related transactions on the decentralized exchanges exceed 250 million dollars in daily trading volume in January 2026.
Prior studies on Solana have examined system performance, smart-contract security, and specific on-chain phenomena. However, we still lack a systematic understanding of what Solana bots implement in practice and how these implementations manifest as observable on-chain execution fingerprints.
To address this gap, we performed a large-scale empirical study of Solana bots from two complementary views: (i) 586 bot repositories collected from GitHub, and (ii) 200 bot addresses on Solana, with over 44 million on-chain transactions. 
Our study derives an implementation-grounded taxonomy of Solana bots comprising 15 categories grouped into five domains (e.g., Trading Operations, MEV, and On-chain Analytics), identifies a largely shared five-stage operational pipeline manifested in bot implementations, and uncovers systematic variation in on-chain trading behaviors of Solana bots across diverse trading platforms and assets. 
Based on our findings, we highlight future research directions, and provide recommendations for building and operating bots on the Solana blockchain.

\end{abstract}

\vspace{-20pt}




\maketitle

\vspace{-5pt} 
\section{Introduction}

Solana is an emerging blockchain platform designed to offer high throughput with low transaction fees~\cite{yakovenko2018solana}, making it an increasingly popular infrastructure for Web3 applications, including decentralized finance (DeFi) and non-fungible tokens  (NFTs)~\cite{hendler2009web3,chen2020blockchain,chohan2021non}. 
Solana adopts a program-account design~\cite{solana_doc} that separates executable logic (programs, i.e., smart contracts on Solana) from mutable state (stored in on-chain accounts), enabling the runtime to execute non-conflicting transactions in parallel.
Together with Proof of History and the Tower BFT consensus protocol~\cite{solana_consensus}, these design choices contribute to the high throughput and low transaction fees of Solana.
As of May 2025, Solana had processed over 405 billion transactions, and supported more than 29.7 million fee-paying accounts~\cite{solana_official}, with total value locked in DeFi exceeding 9 billion USD~\cite{defillama}. 

On the flip side, low transaction fees of Solana make it inexpensive to submit transactions at scale, exposing the network to bot spamming and related financial exploitation~\cite{bots1, bots2}. 
In particular, the daily trading volume of decentralized exchange (DEX) attributed to trading bots has exceeded 250 million USD in January 2026~\cite{bots_tx_2}, and the revenue of these bots peaked at nearly 1.5 million USD in a single day in early 2024~\cite{bots_tx}.
Bots are typically off-chain software systems that interact with the blockchain by constructing and submitting transactions~\cite{niedermayer2024bots}.
Accordingly, a bot rarely corresponds to a single on-chain entity; instead, it manifests through the set of accounts it controls and the on-chain records of the transactions it submits.
Previous studies on Solana span system performance, smart contract security, and measurement of on-chain phenomena. 
Researchers have evaluated the scalability and transaction throughput~\cite{duffy2021IoT, pierro2022scalability}, and proposed vulnerability detection and testing techniques to improve smart contract security~\cite{vrust, smolka2023fuzz}. 
More recently, large-scale empirical studies have measured diverse on-chain phenomena on Solana, including failed transactions~\cite{zheng2025does}, sandwiching attacks~\cite{gerzon2025quantifying}, meme coin markets~\cite{li2025trust}, and rug pulls~\cite{alhaidari2025solrpds}. 
However, Solana bots remain underexplored as software artifacts from a software engineering perspective. We still lack a systematic understanding of what they implement in practice through their codebases, and \hl{the extent to which these implementations correspond to} the on-chain fingerprints left by the transactions they submit.
Such an understanding \hl{provides a complementary characterization of Solana bots from both implementation and execution perspectives}, and provides practical insights into bot design choices and maintenance concerns in a competitive execution environment.

To address this gap, we conducted a large-scale empirical study of Solana bots to systematically characterize their functional categories, implementations in open-source codebases, and transaction-level fingerprints they leave on-blockchain. 
We curated a dataset comprising 586 Solana bot repositories on GitHub and 200 bot accounts from two widely used Solana bot services, with 44,118,825 on-chain transactions submitted by these accounts.
Below, we present our research questions and highlight the key findings:

\noindent \textbf{RQ1: What functionality taxonomy emerges from open-source Solana bot implementations?}
Prior studies often characterize blockchain bots (i.e., on Ethereum and Binance Smart Chain) through their on-chain behaviors or self-reported descriptions~\cite{cernera2023token,cernera2023ready,niedermayer2024detecting,li2023towards}, leaving a limited understanding of what functionalities are actually implemented by bots in open-source repositories.
To establish an implementation-grounded view, we analyze 586 public Solana bot repositories on GitHub, and derive a functionality taxonomy spanning 5 domains and 15 categories, covering \emph{Trading Operations}, \emph{MEV}~\cite{MEV}, \emph{On-chain Analytics}, and \emph{Tooling and Infrastructure}.
The distribution of repositories across the resulting taxonomy indicates that open-source Solana bots are dominated by trading automation and MEV-oriented execution, with analytics and tooling/infrastructure categories constituting a smaller but diverse long tail.

\noindent \textbf{RQ2: How are Solana bots implemented in open-source repositories?}
To complement the functionality taxonomy with an engineering perspective, we explore how Solana bot capabilities are realized in code across bot categories, by analyzing (i) code-level building blocks, (ii) recurring pipeline among these blocks, and (iii) the third-party dependencies underpinning such implementations.
We observe a pronounced long tail in building-block coverage: 90\% of the 2,044 building blocks appear in no more than 3.1\% of repositories, and per-repository counts differ significantly across bot categories.
Despite this heterogeneity, most repositories align with a shared backbone that can be abstracted into a five-stage canonical pipeline spanning \emph{Setup}, \emph{Observation and Acquisition}, \emph{Analytics}, \emph{Planning and Decision}, and \emph{Execution and Reporting}, although the \emph{Planning and Decision} stage is simplified or omitted in certain bot categories.
Dependency analysis further reveals heavy reliance on core Solana SDKs (e.g., \texttt{@solana/web3.js} in 85.9\% of JavaScript/TypeScript repositories) and substantial technical lag, with over 30\% of dependencies more than one year behind the latest versions.

\noindent\textbf{RQ3: How do Solana bots manifest distinctive execution fingerprints on-chain?}
While the implementations and runtime of Solana bots reside off-chain, they are only observable on-chain through the accounts they control and the transactions they submit.
Thus, we characterize the 200 bot-associated addresses in our dataset by extracting on-chain execution fingerprints from their transaction traces, capturing transaction-level features including submission intensity, execution effectiveness, cost, and trading assets.
We observe four distinct execution clusters among the 200 bot-associated addresses, separated primarily along two axes: transaction submission intensity and execution effectiveness.
Three clusters (56 addresses) exhibit execution characteristics consistent with \textit{MEV} bots. Across the three clusters, WSOL~\cite{wsol} serves as the dominant pivot token, accounting for 70.9\%--99.9\% of their transactions; however, the three clusters diverge in the trading venues they engage with. 
Notably, the \textit{MEV} bots whose transactions invoke proprietary AMMs~\cite{prop_amm} (e.g., HumidiFi) realize positive profit (62.3\%) at three times the rate of those that do not, suggesting that interaction with proprietary AMMs may \hl{be associated with more favorable execution outcomes} for MEV bots.
The remaining cluster (102 addresses) exhibits characteristics consistent with the \textit{Trading Operations} bots, with transaction activity concentrated on the Pump.fun platform (80.9\%).

Based on our findings, we discuss the implications, such as recurring operational pipeline across bot implementations,  substantial technical lag of third-party dependencies, and factors that affect the profitability of Solana bots.
We also provide practical recommendations for bot design and maintenance, and highlight several research avenues.
Our work makes the following contributions: (1) We compile two datasets relevant to Solana bots: 586 bot repositories on GitHub, and 200 bot-associated on-chain addresses, with 44,118,825 transactions. (2) We perform the first empirical study of Solana bots to characterize their functional categories, implementations in open-source codebases, and transaction-level fingerprints they leave on-blockchain. (3) We provide practical implications and outline future avenues for research.

\vspace{-10pt}
\section{Background}

Solana bots submit transactions composed of instructions that invoke on-chain programs, exposing behavioral fingerprints across multiple dimensions, such as in program usage and fees.

\noindent\textbf{Accounts and Programs}. Solana adopts an account-based model, where \emph{accounts}, identified by a 32-byte public key (address), store persistent state and executable logic~\cite{solana_doc}.
A \emph{program} in Solana is an executable account that contains program code. 
Solana programs are stateless: 
they read and update state stored in (non-executable) data accounts passed in at execution time~\cite{solana_programs}.

\noindent\textbf{Transactions and Signers.} 
A Solana transaction specifies a set of accounts and a sequence of instructions to be executed by on-chain programs~\cite{instruction}. 
To authorize state changes, a transaction must be signed by one or more \emph{signer accounts} (i.e., accounts whose private keys approve the requested operations), and the corresponding signatures are included in the transaction message.
A \textit{finalized} transaction is \textbf{\emph{successful}} if all instructions execute without error and \textbf{\emph{failed}} otherwise. Both outcomes incur a base fee of 5,000 lamports per signature (0.000005~SOL) and may include an optional per-compute-unit priority fee for higher scheduling priority~\cite{helius_priority_fees}.

\noindent\textbf{MEV.}
MEV refers to the maximum profit obtained by strategically including, excluding, or ordering transactions~\cite{daian2020flashboys}.
MEV opportunities often arise from order-dependent execution on decentralized exchanges (DEXs) and transient price discrepancies of digital assets such as cryptocurrencies across venues (e.g., between DEXs and centralized exchanges, CEXs).
For instance, on mempool-based chains, traders can scan pending price-moving transactions and position their own transactions to capture the resulting price movement.
Unlike Ethereum, Solana does not expose an publicly visible mempool. Instead, Solana adopts mempool-less transaction forwarding (i.e., Gulf Stream~\cite{solana2019gulfstream}), whereby transactions are proactively forwarded to upcoming slot leaders.
Accordingly, MEV on Solana is shaped by latency advantages in reaching leaders and by specialized submission and ordering infrastructure. For instance, Jito provides low-latency transaction submission and bundle-based mechanisms~\cite{jito2025docs}.

\vspace{-5pt}
\section{Dataset}\label{sec:data}

\noindent\textbf{GitHub Repositories of Solana Bots.}
We collected 628 Solana bot via a two-round keyword search using the GitHub GraphQL API, by following a systematic dataset construction process~\cite{wohlin2012experimentation}. 
Specifically, we queried repositories created between January 2021 and June 2025 using the GitHub GraphQL API by matching keywords in repository metadata (e.g., name, description, and topics) and README content. 
The initial query ``\textit{Solana bot}'' returned 7,760 candidates. 
Following prior work~\cite{aghili2023studying}, we applied automated repository-level filters to exclude non-code, link-curation/aggregator, low-engagement, and Solidity repositories, leaving 824 candidates. 
Two authors independently screened each candidate for inclusion using a shared guideline: we required evidence of (i) Solana-specific interaction from README, dependencies, or source code, and (ii) automation-oriented execution (e.g., monitoring loops or automated actions), yielding 536 Solana bot repositories.
Moreover, to reduce keyword bias, we derived three additional query keywords from repository topics observed in the 536 repositories (``\textit{Solana bundler}'', ``\textit{Solana automation}'', and ``\textit{Solana agent}''), which returned 3,046 candidates. 
After applying the same filtering and screening procedures, we obtained 92 additional Solana bot repositories.
We then excluded 42 repositories that could not be processed by an open-source parser generator tool \textsc{tree-sitter}~\cite{tree_sitter_github}
for further investigation, due to missing source code or unsupported languages, resulting in a final dataset of 586 repositories.
Among the 586 Solana bot repositories, TypeScript accounts for the largest share (48.2\%), followed by Python (17.9\%), JavaScript (15.5\%), Rust (14.9\%), and Go (1.9\%). Repository popularity varies widely, with a median of 18 stars (mean: 62, min: 2, max: 2,220, std.: 159), and 7 forks (mean: 27, min: 2, max: 1,019, std.: 70).

\noindent\textbf{On-chain Addresses of Solana Bots.}
\hl{\blackcircle{1}~\textit{\textbf{Baseline Window.}}} To collect on-chain addresses of Solana bots, we started from two widely used bot services on Solana, \emph{Trojan}~\cite{trojanonsolana}
and \emph{SolanaMevBot}~\cite{solanamevbot_docs}.
Trojan is a Telegram-based bot service that supports automated trading, sniping, and copy trading, and is listed among the top services in the DEX trading-bot leaderboard on Dune Analytics~\cite{dune_dex_trading_bot_wars}.
SolanaMevBot is listed among the top services in the arbitrage leaderboard on circular.bot~\cite{circular_bots}.
Some high-ranked services (e.g., \emph{NotArb} on circular.bot) do not provide publicly accessible on-chain bot addresses, and are therefore excluded from our analysis.
For each selected bot service, we collected the top-100 bot addresses on Solana reported by the corresponding leaderboard. For each address, we leveraged the Dune SQL engine~\cite{dune_analytics}
to retrieve its on-chain transaction records over a one-month window from October 1, 2025, to November 1, 2025, including transaction signatures, execution status, invoked program IDs, transaction fees, and token balance changes, as well as-level traces.
The transaction volume per address during our one-month observation window varies substantially across bot services. Bot addresses from  SolanaMevBot range from 5 to 6,742,370 transactions (median: 2,121,  mean: 439,061). In contrast, bot addresses from Trojan range from 64 to 28,274 transactions (median: 1,366, mean: 2,127).
\hl{\blackcircle{2}~\textit{\textbf{Validation Window.}} We constructed an additional validation dataset over a one-month window from February 1, 2026, to March 1, 2026. Specifically, we collected the top-100 bot addresses reported for SolanaMevBot on circular.bot, as well as the top-100 addresses reported for Axiom on Dune Analytics~\cite{axiom_dune} during this window. Axiom is a browser-based Solana trading-bot service that was listed among the top services in the DEX trading-bot leaderboard on Dune Analytics during this period. For each address, we applied the same Dune SQL-based data collection procedure to retrieve its on-chain transaction records. See Appendix~\ref{sec:validation-window} for details.
}

\vspace{-5pt}
\section{RQ1: Taxonomy}\label{sec:RQ1}
To answer RQ1, we constructed a taxonomy of Solana bot repositories to systematically capture the functionalities implemented by these bots.
We use Solana bot repositories on GitHub as inputs because they provide inspectable and versioned implementations of bot functionalities, allowing the taxonomy to be grounded in what is actually implemented rather than what is merely described.
\vspace{-10pt}
\subsection{Methodology}  
We applied an LLM-assisted open card sorting process to construct a taxonomy of functionalities implemented in Solana bot repositories on GitHub, because manually analyzing repository codebases is impractical given their scale and heterogeneity~\cite{dhulshette2025hierarchical,oskooei2025repository}. 
Specifically, we decomposed the process into four steps and used an LLM (DeepSeek v3.2) as an auxiliary tool for code summarization and higher-level semantic abstraction. The full set of prompts is provided in the replication package.

\noindent\blackcircle{1}\textbf{ Function-Level Card Preparation.}
We used \textsc{tree-sitter} to extract all functions from each repository. For each extracted function, we provided its source code as input to the LLM, and obtained a concise natural language summary. These function-level summaries serve as cards in the subsequent open card sorting process.
\noindent\blackcircle{2}\textbf{ Repository-Level Summarization.}
For each repository, we aggregated its set of function-level cards, and prompted the LLM to generate descriptive functionality tags, along with the proportion of the function-level cards in the repository associated with each tag. These tags summarize salient functional aspects of the repository as a whole and facilitate higher-level grouping and taxonomy construction. As a result, we generated an average of 6.3 tags per repository (min = 5, median = 6, max = 9, std. = 1). 
\noindent\blackcircle{3}\textbf{ Taxonomy Construction.}
To construct the taxonomy, we first consolidated synonymous tags across repositories (e.g., \textit{solana-sniper-bot} and \textit{sniping-bot}), 
yielding a canonicalized tag set (N = 372). \hl{Two of the authors then independently performed open card sorting on the canonicalized tag set.
The two initial taxonomies converged at the domain level, differing in category granularity (e.g., one taxonomy merging semantically related MEV-oriented categories that the other kept separate) and category naming (e.g., differently worded labels covering the same underlying tag groups). The two authors reconciled these differences into a finalized taxonomy of 15 categories. Using finalized taxonomy, the two authors independently assigned the 372 canonicalized tags to one of the 15 categories; a tag was counted as agreed upon if both authors assigned it to the same category. 
These tag-level assignments achieved substantial inter-rater agreement (Cohen’s $\kappa = 0.78$).
Disagreements were resolved through discussion between the two authors; unresolved cases were independently reviewed by a third author, and the majority decision was adopted as the final assignment.
}
\noindent\blackcircle{4}\textbf{ Repository Categorization.}
We applied the taxonomy by mapping each repository to one or more taxonomy categories based on its repository-level tags. 
As a result, each repository is associated with 1 to 7 categories, while 16 repositories cannot be mapped to any category.
As a sanity check, we randomly sampled 100 out of 586 repositories, \hl{allocating the sample across the 15 categories in proportion to the square root of each category's size.
The strategy oversamples smaller categories compared to simple proportional allocation to balance estimation reliability across strata of varying sizes~\cite{cochran1977sampling}.
Two authors independently reviewed each sampled repository's README and source code, and judged whether the repository's dominant category accurately reflected a primary functionality of the repository, with substantial inter-rater agreement (Cohen's $\kappa = 0.81$).}
Note that we define the dominant category of a repository as the category of its highest-weight tag (i.e., the tag covering the largest proportion of function-level cards).
\hl{
Disagreements were resolved through discussion between the two authors until consensus was reached. 
91 of the 100 sampled repositories were confirmed to have a dominant category that accurately reflected a primary functionality of the repository, with per-category accuracy ranging from 75\% to 100\%. See Table~\ref{tab:repo_validation} in the Appendix for details.} We further grouped repositories by dominant category for subsequent analysis.

\vspace{-10pt}
\subsection{Results}

\begin{table*}[t]
\centering
\scriptsize
\caption{Taxonomy of Solana Bot Repositories (Repos) on GitHub.}

\label{tab:taxonomy}
\rowcolors{2}{gray!10}{white}
\renewcommand{\arraystretch}{1.05}
\resizebox{0.85\linewidth}{!}{
\begin{tabular}{p{4cm} p{8.2cm} p{5cm}}
\toprule
\textbf{Category (\# Repos)} & \textbf{Example Tags} & \textbf{Representative Repo} \\
\hline

\multicolumn{3}{l}{\cellcolor{lightgray}\textbf{\faRobot~Trading Operations}}\\
\textbf{Order Execution and Management (137)}
&
\textit{trading-bot}, \textit{automated-trading}, \textit{transaction-execution}, \textit{limit-orders}, \textit{telegram-trading-bot}
&
\path{nmweaver/soltrade} (342~\tiny \faStar)
\\

\textbf{Liquidity Provision and Yield Farming (12)}
&
\textit{liquidity-management}, \textit{bonding-curve}, \textit{liquidity-bot}, \textit{automated-market-maker}, \textit{liquidity-pools}
&
\path{edwin-finance/meteora-liquidity-rebalancer} (4~\tiny \faStar) 
\\

\textbf{Copy Trading (33)}
&
\textit{copy-trading}, \textit{copy-trading-bot}
&
\path{cryptole0/Copy-Trading-Bot-Rust} (96~\tiny \faStar)
\\

\hline
\multicolumn{3}{l}{\cellcolor{lightgray}\faRobot~\textbf{MEV}}\\
\textbf{Arbitrage (47)}
&
\textit{arbitrage-bot}, \textit{flash-loan-arbitrage}, \textit{cross-pool-arbitrage}, \textit{cross-exchange-arbitrage}, \textit{dex-cex-arbitrage}
&
\path{0xNineteen/solana-arbitrage-bot} (799~\tiny \faStar)
\\

\textbf{Sniping (134)}
&
\textit{sniper-bot}, \textit{pumpfun-bot}, \textit{pumpfun-trading-bot}, \textit{memecoin-trading}, \textit{token-sniper}
&
\path{warp-id/solana-trading-bot} (2,220~\tiny \faStar)
\\

\textbf{Transaction Ordering Exploitation (41)}
&
\textit{mev-bot}, \textit{jito-bundle}, \textit{transaction-bundling}, \textit{pumpfun-bundler}, \textit{sandwich-attack}
&
\path{jito-labs/mev-bot} (1,150~\tiny \faStar)
\\

\hline
\multicolumn{3}{l}{\cellcolor{lightgray}\textbf{\faRobot~Market Manipulation}}\\
\textbf{Wash Trading (21)}
&
\textit{volume-bot}, \textit{volume-simulation}
&
\path{web3batman/Raydium-Volume-Bot} (67~{\tiny\faStar})
\\

\hline
\multicolumn{3}{l}{\cellcolor{lightgray}\textbf{\faRobot~On-chain Analytics}}\\
\textbf{Monitoring and Alerting (26)}
&
\textit{real-time-alerts}, \textit{token-monitoring}, \textit{real-time-monitoring}, \textit{wallet-tracking}, \textit{transaction-monitoring}
&
\path{FriedDev/solana-rug-checker} (29~{\tiny\faStar})
\\

\textbf{Portfolio and Risk Management (14)}
&
\textit{risk-management}, \textit{risk-analysis}, \textit{portfolio-management}, \textit{simulation}, \textit{data-analysis}
&
\path{henrytirla/Solana-PNL-Bot} (66~{\tiny\faStar})
\\

\hline
\multicolumn{3}{l}{\cellcolor{lightgray}\textbf{\faCogs~Tooling and Infrastructure}}\\
\textbf{Token Issuance and Administration (8)}
&
\textit{token-launchpad}, \textit{token-creation}, \textit{token-launch}, \textit{token-management}, \textit{spl-token}
&
\path{0xNevo/Solana_Token_Freezer} (14~{\tiny\faStar})
\\

\textbf{NFT Minting and Marketplace Trading (14)}
&
\textit{nft-sales-bot}, \textit{nft-tools}, \textit{nft-minting-bot}, \textit{nft}, \textit{nft-bot}
&
\path{theskeletoncrew/air-support} (178~{\tiny\faStar})
\\

\textbf{DEX Swap Routing and Integration (30)}
&
\textit{raydium-integration}, \textit{jupiter-aggregator}, \textit{dex-integration}, \textit{jupiter-swap}, \textit{dex-aggregator}
&
\path{YZYLAB/solana-swap} (130~{\tiny\faStar})
\\

\textbf{Wallet Management (6)}
&
\textit{wallet-management}, \textit{multi-wallet}, \textit{token-transfer}, \textit{account-management}, \textit{keypair-management}
&
\path{LeaderMalang/Solana-Sweeper-Bot} (9~{\tiny\faStar})
\\

\textbf{On-chain Data Pipeline (10)}
&
\textit{rpc-client}, \textit{data-aggregation}, \textit{birdeye-api}, \textit{geyser-client}, \textit{helius-webhook}
&
\path{weeaa/goyser} (66~{\tiny\faStar})
\\

\textbf{Messaging Integration and Alerts (37)}
&
\textit{telegram-bot}, \textit{twitter-integration}, \textit{discord-bot}, \textit{chat-interface}, \textit{discord-notifications}
&
\path{KingJiongEN/DegentGroup} (65~{\tiny\faStar})
\\

\bottomrule
\end{tabular}
}
\vspace{-10pt}
\end{table*}

Table~\ref{tab:taxonomy} presents the taxonomy of Solana bot functionalities derived from the open card sorting process based on our dataset of Solana bot repositories on GitHub. 
The taxonomy summarizes how Solana bots group around recurring functional themes, and organizes the observed repository functionalities of Solana bots into 5 higher-level domains and 15 categories, with each category accompanied by representative tags drawn from the repositories assigned to that category (the \textit{Tags} column), and a representative repository for the category. 
For a detailed description of each category, please refer to Table~\ref{tab:taxonomy_detailed} in the Appendix~\cite{appendix}.

We further conducted a qualitative analysis of the most-starred repository in each of the top three categories (ranked by repository count), focusing on their core functionalities. Below is a representative repository (additional examples are provided in the Appendix): 

\faRobot~The \texttt{0xNineteen/solana-arbitrage-bot}, written in Rust and TypeScript, exemplifies the \emph{Arbitrage} category, which combines (1) an off-chain client that searches for arbitrage opportunities and assembles transactions, and (2) an on-chain program that enforces atomic execution and \emph{profit-or-revert} semantics of transactions. 
Specifically, the off-chain client searches for profitable swap paths across multiple DEX liquidity pools (e.g., Orca, Mercurial, and Saber), and assembles the corresponding swap instructions into a single transaction. The on-chain program validates the profitability of the transaction by tracking realized balances across hops, and reverts the entire transaction unless the final balance of the source token exceeds its initial balance.
Moreover, we also observed multiple heterogeneous repositories whose functionalities span multiple taxonomy categories. 
Taking the \texttt{sendaifun/solana-agent-kit} repository as an example, it combines bot-like autonomous execution with tooling and infrastructure support, spanning the \textit{trading operations} (\faRobot) and \textit{tooling and infrastructure} (\faCogs) categories.
On the one hand, it supports bot-like automation of on-chain trading-related actions (e.g., token swaps and other DeFi interactions). 
On the other hand, it provides a broad range of infrastructure and utility capabilities beyond trading, such as wallet management and NFT-related operations.
Notably, it integrates with agent runtimes (e.g., LangChain and Vercel AI) by exposing on-chain actions through a unified interface, enabling agent planning modules to invoke both trading and infrastructure operations.
\vspace{-5pt}
\rqbox{
\textbf{Finding 1:} 
We identify a taxonomy of Solana bot functionalities comprising 5 higher-level domains and 15 categories. Open-source Solana bot implementations emphasize automating order placement and reacting to time-sensitive signals.
}\vspace{-10pt}

\section{RQ2: Implementation}\label{sec:RQ2}
To address RQ2, we explore GitHub repositories of Solana bots from an implementation perspective, focusing on the code-level building blocks of Solana bots and how these blocks are composed into typical bot pipelines. 
We also investigate the usage of third-party libraries, which reflect how bot pipeline components are implemented in practice.

\subsection{Methodology}

\subsubsection{Code-Level Building Block Identification}
We used the 78,300 functions extracted from the repositories in our dataset via \textsc{tree-sitter}, and grouped them into code-level building blocks based on lexical and semantic similarity, by following the steps below.

\noindent\blackcircle{1}\textbf{ Function Embedding.} For each function, we preprocessed its code by removing comments and normalizing literals (e.g, numbers and strings) and identifiers (e.g., variable and parameter names). We then concatenated the function name and the normalized function body as the input, and used the Qwen3-Embedding-0.6B model~\cite{qwen3embedding} 
to obtain an embedding vector for the function, with the output dimension of 1,024.
Qwen3-Embedding-0.6B is a lightweight embedding model that excels at code retrieval and text clustering on benchmarks, while offering a strong balance between effectiveness and efficiency for large-scale embedding scenarios~\cite{mu2026hearing,liu2026detection}, making it well-suited for large-scale function embedding.

\noindent\blackcircle{2}\textbf{ Function Clustering.} 
Based on the generated function embeddings, we clustered functions using a UMAP-HDBSCAN pipeline.
UMAP~\cite{mcinnes2018umap} is a nonlinear dimensionality reduction method that projects high-dimensional embeddings into a lower-dimensional space, while largely preserving local neighborhood relationships.
HDBSCAN~\cite{campello2013density} is a non-parametric clustering algorithm, which has demonstrated its efficacy for code clustering in recent studies (e.g., ~\cite{liang2023needle,ma2025surviving,chen2025my}).
In the UMAP-HDBSCAN pipeline, HDBSCAN leverages the local neighborhood structure preserved in the UMAP-reduced space to identify coherent clusters without predefining the number of clusters, which makes the pipeline well-suited to function-level embeddings that often exhibit uneven distributions and outliers.
In particular, we first applied UMAP to learn a lower-dimensional manifold representation of the function embeddings, and then ran HDBSCAN in this space to discover density-based clusters and label outliers as noise.
Moreover, we tuned the hyperparameters of the UMAP-HDBSCAN pipeline via grid search to maximize the Silhouette score while constraining the fraction of functions labeled as noise. Specifically, we tuned \emph{n\_neighbors} and \emph{n\_components} of UMAP as well as \emph{min\_cluster\_size} and \emph{min\_samples} of HDBSCAN, and set them to 10, 10, 5, and 5, respectively.
As a result, we identified 5,007 function clusters, with a Silhouette score of 0.82, suggesting well-separated clusters.
\noindent\blackcircle{3}\textbf{ Cluster Labeling and Consolidation.} For each function cluster, we aggregated the function-level summaries (generated in RQ1) of the functions in the cluster, and used an LLM (DeepSeek-v3.2) to generate one label in a constrained \texttt{[object]-[operation]} format (e.g., \texttt{token-swap}). To improve reproducibility, we fixed the model version and decoding configuration (e.g., \emph{temperature} = 0) and recorded the prompts and generated labels.
To validate labeling quality, \hl{we randomly sampled 95 clusters from the 5,007 function clusters, which provides a margin of error of approximately 10\% at the 95\% confidence level. Two of the authors independently reviewed the top-10 representative functions in each sampled cluster, ranked by HDBSCAN membership probability, and judged whether the generated label captured the functionality commonly observed across the 10 representative functions, based on manual inspection of each function's source code. Cohen's kappa over these independent judgments was $\kappa = 0.71$, indicating substantial agreement. Disagreements were resolved through discussion between the two authors until consensus was reached. After reconciliation, 84 of the 95 sampled clusters were confirmed to have labels that accurately described their functionality.}
Finally, we merged clusters with semantically overlapping labels into code-level building blocks (e.g., \texttt{transaction-construction} vs \texttt{transaction-creation}), yielding 2,003 building blocks in total.

\vspace{-10pt}
\subsubsection{Bot Pipeline Derivation}
We derived bot pipelines that summarized how the identified code-level building blocks are composed in practice across bot categories. Following a bottom-up hierarchical abstraction approach inspired by prior work~\cite{maqbool2007hierarchical}, we modeled a pipeline as an ordered sequence of stages, where each stage grouped building blocks that served a coherent architectural responsibility.
Specifically, for each bot category, we selected multiple representative repositories, and inspected the building blocks in each repository. We then grouped these building blocks into pipeline stages primarily based on architectural responsibilities reflected by repository modules or components, and used their frequent co-occurrence within repositories of the same category as supporting evidence. We further consulted call or usage relations of building blocks when available to infer and validate plausible stage ordering.

\subsubsection{Third-Party Library Usage Characterization} 
We first extracted the dependency entries from the language-specific dependency manifests of bot-related repositories in our dataset, i.e.,  \path{package.json}, \path{requirements.txt}, \path{Cargo.toml}, and \path{go.mod} for repositories written in JavaScript/TypeScript, Python,  Rust, and Go, respectively. Each dependency entry specifies a library (by name) and, when provided, its version constraint.
Next, to restrict our analysis to third-party dependencies, we excluded language-provided modules and packages (i.e., built-in modules or standard libraries) from the extracted library lists. 
Specifically, we obtained the built-in modules and standard libraries from official runtime APIs and toolchain commands, including \path{module.builtinModules} for JavaScript/TypeScript, \path{sys.stdlib_module_names} for Python, and \texttt{go list std} for Go. For Rust, we removed compiler-provided standard-library crates, e.g., \texttt{std}, \texttt{core}, \texttt{alloc}, and \texttt{proc\_macro}.
As a result, we obtained 15,524 dependency entries spanning 3,194 distinct third-party libraries.
The average number of repositories using each third-party library is 4.86 (median = 1, min = 1, max = 328, and std = 15), indicating a highly right-skewed distribution.

Moreover, we characterized the up-to-dateness of third-party library usage in Solana bot repositories by measuring the \emph{technical lag}~\cite{he2023automating,cogo2019empirical} of each dependency entry. For each dependency entry, we resolved an \textit{effective version} as the latest release that satisfies the declared version constraint, and computed \textit{technical lag} as the time gap between the release date of this effective version and the release date of the latest available release of the same library.

\subsection{Results}
\subsubsection{Code-Level Building Blocks in Bot Repositories.}

\begin{figure}[t]
    \centering
    \begin{subfigure}[t]{0.35\linewidth}
        \centering
        \includegraphics[width=\linewidth, trim={0 2mm 0 0.5mm},clip]{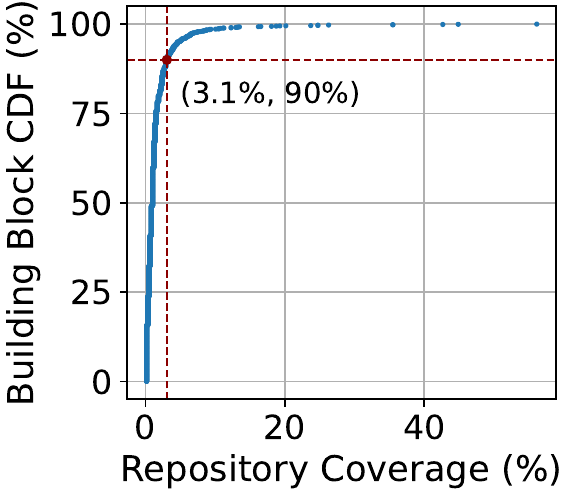}
        \caption{Overall}
        \vspace{-5pt}
        \label{fig:bb_cdf_overall}
    \end{subfigure}
    \hfill
    \begin{subfigure}[t]{0.35\linewidth}
        \centering
        \includegraphics[width=\linewidth, trim={0 2mm 0 0.5mm},clip]{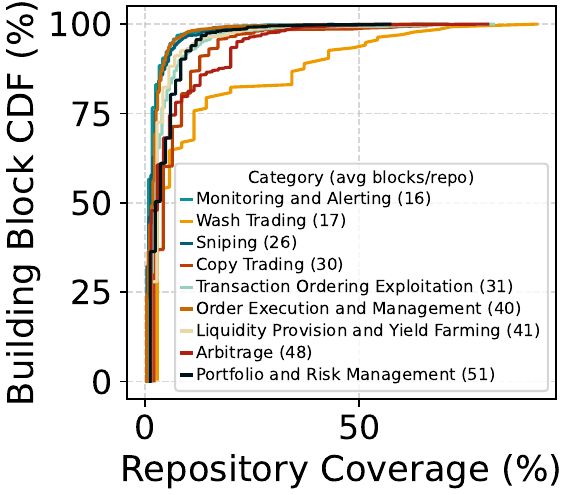}
        \caption{By bot category}
        \vspace{-5pt}
        \label{fig:bb_cdf_category}
    \end{subfigure}
    \caption{Repository coverage   distribution per building block.}
    \label{fig:bb_cdf}
    \vspace{-10pt}
\end{figure}

Figure~\ref{fig:bb_cdf_overall} presents the cumulative distribution of repository coverage across building blocks.
Overall, the repository coverage of building blocks follows a pronounced long-tail distribution, as shown in Figure~\ref{fig:bb_cdf_overall}. Specifically, 90\% of building blocks appear in no more than 3.1\% of Solana bot repositories, indicating that most building blocks have very limited repository coverage. In contrast, a small minority of building blocks attains substantially higher coverage across bot repositories, forming a clear head–tail pattern.
Table~\ref{tab:building_blocks_repo} makes the head of the distribution explicit by listing the ten most prevalent building blocks ranked by repository coverage. The most prevalent building block, \path{trade-execution}, appears in 58.5\% of repositories, followed by \path{token_data_fetching} (46.9\%) and \path{transaction-construction} (45.2\%). Even the tenth-ranked building block is present in 21.8\% of repositories, indicating that no building block approaches near-universal presence.
The head concentrates on two capability families, (1) data acquisition (e.g., \path{token} / \path{pool} / \path{market} / \path{wallet}\path{_data-fetching}), and (2) the transaction pipeline (\path{transaction-construction} / \path{submission}, and \path{trade-execution}), revealing a common ``observe--trade'' backbone across bot repositories. The presence of \texttt{bundle-submission} among the most prevalent building blocks further indicates explicit support for batched/bundled submission beyond standard transaction submission.

\begin{table}[t]
\centering
\caption{Top-10 building blocks by repository coverage.}
\label{tab:building_blocks_repo}
\small
\resizebox{\linewidth}{!}{
\begin{tabular}{lp{8cm} p{3.1cm}}
\toprule
\textbf{Building Block} & \textbf{Description} & \textbf{Repo Coverage (n, \%)} \\
\midrule
\texttt{trade-execution} & Executes token swap or order operations.   & 333 (58.5\%) \\
\texttt{token\_data-fetching} & Retrieves on-chain token state, including price, supply, and identifiers. & 267 (46.9\%) \\
\texttt{transaction-construction} & Constructs transaction or instruction objects for on-chain operations.  & 257 (45.2\%) \\
\texttt{transaction-submission} & Submits signed transactions to Solana network for on-chain execution.  & 215 (37.8\%) \\
\texttt{pool\_data-fetching} & Retrieves the configuration and state of a liquidity pool. & 210 (36.9\%) \\
\texttt{token\_account\_data-fetching} & Retrieves token account state, including balance, mint, and ownership.  & 157 (27.6\%) \\
\texttt{market\_data-fetching} & Retrieves aggregated market-level data, such as price feeds, trading volume, or cross-venue liquidity metrics. & 150 (26.4\%) \\
\texttt{client-creation} & Initializes RPC/API clients for interacting with Solana nodes or programs.  & 148 (26.0\%) \\
\texttt{bundle-submission} & Bundles and submits multiple transactions. & 146 (25.7\%) \\
\texttt{wallet\_data-fetching} & Retrieves wallet-level state (e.g., token balances). & 124 (21.8\%) \\
\bottomrule
\end{tabular}
} 
\vspace{-15pt}
\end{table}

As illustrated in Figure~\ref{fig:bb_cdf_category}, the average number of building blocks per repository varies across bot categories, ranging from 16 for the \emph{Monitoring and Alerting} bots to 51 for the \emph{Portfolio and Risk Management} bots.
We applied the Kruskal--Wallis test and observed a statistically significant difference in building-block counts across bot categories ($H = 43.172$, $p = 8.152 \times 10^{-7}$).
Moreover, the counts of building blocks per repository are decoupled from the degree of sharing in building blocks across repositories, as shown in Figure~\ref{fig:bb_cdf_category}.
Specifically, building-block composition across bot categories exhibits three distinct types, including (1) \textit{\textbf{low-count, core-shared composition}}, e.g., \emph{Wash Trading} bots contain an average of 17 building blocks, with 18\% of the building blocks appearing in more than 30\% of the corresponding repositories; (2) \textit{\textbf{low-count, weakly shared composition}}, e.g., \emph{Monitoring and Alerting} and \emph{Sniping} bots both remain small in size, with 16 and 26 building blocks per repository on average, yet less than 1\% of their building blocks appear in more than 30\% of the corresponding repositories; and (3) \textit{\textbf{high-count, long-tail–dominated composition}}, e.g., \emph{Arbitrage bots} exhibit large block count (48 in average), while over 80\% of their building blocks appear in fewer than 10\% of repositories.
\vspace{-5pt}
\rqbox{
\textbf{Finding 2:} 
The repository coverage of building blocks exhibits a long-tail distribution, suggesting an ``observe-trade'' backbone across bot repositories. The building-block counts per repository differ statistically significantly across bot categories.
}\vspace{-10pt}

\subsubsection{Bot Pipelines}
\begin{figure*}[htbp]
    \centering
    \includegraphics[width=0.8\linewidth, trim={0 6.5mm 0 5mm},clip]{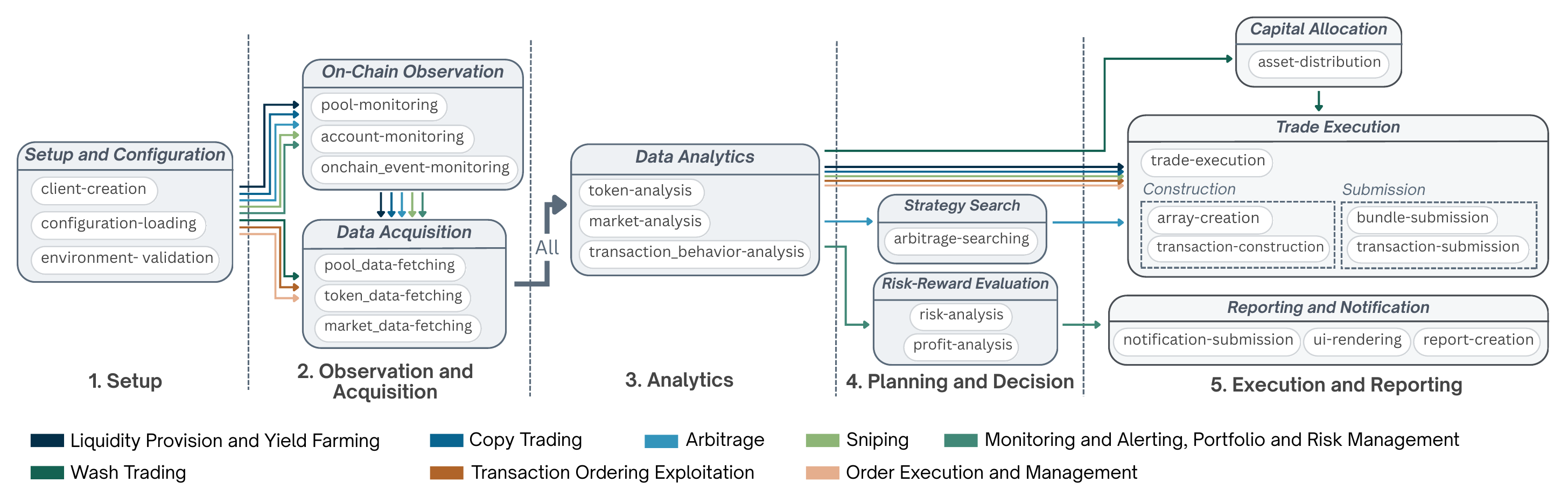}
    \caption{Bot pipeline across categories, organized into five stages.} 
    \label{fig:workflow}
    \vspace{-10pt}
\end{figure*}

Figure~\ref{fig:workflow} summarizes an abstracted, stage-based pipeline view across Solana bot categories, highlighting both shared components and category-level variations of Solana bots.
At a high level, the pipeline of Solana bots is organized into five stages:
(1) \textbf{\textit{Setup}}, which covers client initialization and configuration (e.g., establishing connections to Solana RPC services);
(2)\textbf{ \textit{Observation and Acquisition}}, which includes on-chain monitoring (e.g., liquidity pools, accounts, and events such as new token listings) and data fetching (e.g., token, pool, and market data);
(3) \textbf{\textit{Analytics}}, which covers data processing and analysis (e.g., analyses of tokens, market and transaction behavior);
(4) \textbf{\textit{Planning and Decision}}, which covers pre-execution logic such as strategy search and risk-reward evaluation; and
(5)\textbf{ \textit{Execution and Reporting}}, which includes capital allocation, trade execution, and post-trade reporting/notification.

Several components recur across bot categories, such as the \textbf{\textit{Data Acquisition}}, \textbf{\textit{Data Analytics}}, and \textbf{\textit{Trade Execution}} modules, indicating a common execution backbone of Solana bots. 
We also observe that the \textbf{\textit{Trade Execution}} module is decomposed into transaction construction (e.g., \path{array-creation} and \path{transaction-construction}) and transaction submission, where Solana bot repositories tend to adopt two different submission modes, including direct transaction submission (e.g., \texttt{transaction-submission}) and bundle-based submission (e.g., \texttt{bundle-submission}).

Differences across bot categories primarily arise from which functional components are traversed and how building blocks are composed within stages.
Bots in some categories tend to compose additional components in the \textbf{\textit{Planning and Execution}} stage, such as Strategy Search in \textit{Arbitrage} bots, Risk-Reward Evaluation in \textit{On-chain Analytics} bots.
In contrast, bots in other categories reach \textit{Trade Execution} with fewer intermediate components, often bypassing the \textbf{\textit{Planning and Execution}} stage, and reach \textbf{\textit{Trade Execution}} with fewer intermediate components, e.g., those in the \textit{Order Execution and Management} and \textit{Transaction Ordering Exploitation} categories. 

\vspace{-5pt}
\rqbox{
\textbf{Finding 3:} 
Solana bots follow a five-stage pipeline at a high level.
Category-level variations arise from which functional components are traversed and how building blocks are composed within stages.
}\vspace{-10pt}

\subsubsection{Third-Party Library Usage}
Table~\ref{tab:library_usage} presents the five most frequently used third-party libraries in Solana bot repositories across programming languages.
We made the following observations: 
\noindent(1) \textbf{\textit{Dominance of the Solana client SDK.}} Solana client SDK libraries are widely adopted across languages, with coverage of 85.9\% for \texttt{@solana/web3.js} in JavaScript/TypeScript, 77.6\% for \texttt{solana-sdk} in Rust, 53.4\% for \texttt{solana} in Python, and 80.0\% for \texttt{solana-go} in Go repositories, respectively.
\noindent(2) \textbf{\textit{Frequent adoption of general-purpose libraries.}}
For configuration loading, \texttt{dotenv} appears in 76.4\% of JavaScript/TypeScript repositories, and \texttt{python-dotenv} appears in 50.7\% of Python repositories.
For HTTP communication, the coverage levels are 60.5\% for \texttt{axios} in JavaScript/TypeScript repositories and 75.3\% for \texttt{requests} in Python repositories.
For Base58 encoding and decoding, \texttt{bs58} is adopted by 64.7\% of JavaScript/TypeScript repositories, and \texttt{base58} is adopted by 46.6\% of Python repositories.
\noindent(3) \textbf{\textit{Differences in coverage uniformity across languages.}}
In Rust repositories, the top libraries exhibit relatively similar coverage levels (71\%--78\%):  
77.6\% for \texttt{solana-sdk}, 76.5\% for \texttt{serde} and 71.8\% for \texttt{tokio}.
In contrast, in Python repositories, Solana-related libraries, i.e., \texttt{solana} (53.4\%) and \texttt{solders} (47.9\%), have substantially lower coverage as compared to the HTTP client library \texttt{requests} (75.3\%).

\begin{table}[t]
\centering
\small
\renewcommand{\arraystretch}{0.9}
\caption{Top five third-party libraries frequently used by Solana bot repositories, grouped by programming language.}
\label{tab:library_usage}
\resizebox{0.75\linewidth}{!}{
\begin{tabular}{llrr}
\toprule
\textbf{\makecell{Programming  Language}} & \textbf{Library} & \textbf{\makecell{Repo Coverage (n, \%)}} & \textbf{Technical Lag}\\
\midrule
\multirow{5}{*}{\makecell{JavaScript / \\ TypeScript}}
& \texttt{@solana/web3.js}      & 328 (85.9\%) & 125 days\\
& \texttt{dotenv}              & 292 (76.4\%) & 223 days\\
& \texttt{@solana/spl-token}   & 263 (68.8\%) & 0 days\\
& \texttt{bs58}                & 247 (64.7\%) & 857 days\\
& \texttt{axios}               & 231 (60.5\%) & 0 days\\

\midrule
\multirow{5}{*}{Rust}
& \texttt{solana-sdk}          & 66 (77.6\%) & 310 days\\
& \texttt{serde}               & 65 (76.5\%) & 0 days\\
& \texttt{serde\_json}         & 62 (72.9\%) & 0 days\\
& \texttt{tokio}               & 61 (71.8\%) & 0 days\\
& \texttt{anyhow}              & 56 (65.9\%) & 0 days\\

\midrule
\multirow{5}{*}{Python}
& \texttt{requests}            & 55 (75.3\%) & 446 days\\
& \texttt{solana}              & 39 (53.4\%) & 572 days\\
& \texttt{python-dotenv}       & 37 (50.7\%) & 642 days\\
& \texttt{solders}             & 35 (47.9\%) & 611 days\\
& \texttt{base58}              & 34 (46.6\%) & 0 days\\

\midrule
\multirow{5}{*}{Go}
& \texttt{solana-go}           & 8 (80.0\%)  & 386.5 days\\
& \texttt{binary}              & 6 (60.0\%)  & 446 days\\
& \texttt{godotenv}            & 6 (60.0\%)  & 140 days\\
& \texttt{grpc}                & 6 (60.0\%)  & 679 days\\
& \texttt{testify}             & 5 (50.0\%)  & 584 days\\

\bottomrule
\end{tabular}
}
\vspace{-15pt}
\end{table}

Regarding dependency technical lag, the median lag across all third-party library dependencies in Solana bot repositories is 23 days; more than 30\% of dependencies lag by over one year, indicating substantial dependency staleness for a non-trivial portion of dependencies.
In terms of frequently used third-party libraries, technical lag varies substantially across their dependencies in Solana bot repositories of different programming languages, as shown in the \textit{Technical Lag} column in Table~\ref{tab:library_usage}. In particular, the dependencies of seven top-used libraries have a median technical lag of 0 days, whereas the dependencies of nine have a median technical lag exceeding one year.
Across programming languages, Rust bot repositories show concentrated technical lag primarily in the core SDK (\texttt{solana-sdk}: 310 days), whereas Python bot repositories exhibit more systematic technical lag among their top-used library dependencies.
In particular, four of the top five libraries used by Python bot repositories have a median technical lag exceeding one year, i.e., \texttt{python-dotenv} (642 days), \texttt{solders} (611 days),  \texttt{solana} (572 days), and \texttt{requests} (446 days), reflecting notable dependency staleness in Python bot repositories.

\hl{In addition to programming language, dependency lag also differs significantly across bot categories. Repositories were grouped by their dominant taxonomy category, with repository-level lag summarized as the median across third-party dependencies. The differences are statistically significant (Kruskal--Wallis, H = 48.029, p = $1.3\times {10}^{-5}$). \emph{NFT minting and marketplace trading} bots exhibit the highest median repository-level lag (433.9 days), followed by \emph{arbitrage} bots (391.2 days), whereas several categories, including \emph{copy trading}, \emph{sniping}, and \emph{monitoring and alerting}, have a median lag of 0 days, suggesting that maintenance practices vary substantially across functional roles (see Table~\ref{tab:category_dependency_lag} in the Appendix).}

We further examined trading-venue-specific package dependencies across bot repositories, such as \texttt{@raydium-io/raydium-sdk} for Raydium, \texttt{@jup-ag/core} for Jupiter, \texttt{@orca-so/common-sdk} for Orca, and \texttt{@meteora-ag/dlmm} for Meteora. We observe that Raydium-related dependencies are the most widely adopted, appearing in 195 bot repositories (33.3\%), substantially exceeding Jupiter (40 repositories), Orca (45), Pump.fun (36), and Meteora (30).
Beyond overall adoption, the breadth of venue integration also varies substantially across bot categories. In particular, \textit{Arbitrage} bot repositories most frequently rely on multiple venues for trading, with 24\% depending on two or more venues (average: 0.88), consistent with their need to identify and exploit price discrepancies across venues.
In contrast, \textit{Monitoring and Alerting} bot repositories exhibit much narrower venue integration, with only 5.5\% depending on multiple venues (average: 0.41), reflecting a more specialized, single-venue trading strategy.
\vspace{-5pt}
\rqbox{
\textbf{Finding 4:} 
Solana bot repositories show concentrated dependence on a small set of third-party libraries. More than 30\% of dependencies lag by over one year. Among frequently used third-party libraries, technical lag varies substantially across programming languages and repositories.

}\vspace{-12pt}

\section{RQ3: On-chain Fingerprints}\label{sec:RQ3}

Understanding the on-chain fingerprints of Solana bots requires examining how bot-associated addresses execute transactions under competitive on-chain conditions, e.g., contention for block inclusion and transaction execution.
Per-address transaction traces are often noisy, and the underlying intent of bots as operationalized by the proposed taxonomy in RQ1 could only be indirectly inferred from execution traces.
To mitigate noise and avoid relying on explicit intent labels, we cluster bot addresses with similar execution patterns and characterize on-chain fingerprints at the cluster level.

\subsection{Methodology}
\subsubsection{Address Clustering}
We first represented each bot-associated address as a feature vector capturing its execution characteristics.
In line with prior work~\cite{zheng2025does}, we adopted two intensity features (\emph{transaction frequency} and \emph{transaction volume}), and extended them with three additional features capturing execution effectiveness, cost, and breadth (\emph{transaction success rate}, \emph{transaction fee}, and \emph{asset diversity}).
Specifically, for each address, we measured \emph{transaction frequency} as the mean inter-transaction interval between transactions initiated by the address; \emph{transaction volume} by the total number of initiated transactions and the average number of initiated transactions per block over the analysis window; \emph{transaction success rate} as the fraction of successful initiated transactions, where a transaction was marked as successful if its \texttt{error} field is null and failed otherwise; 
\emph{transaction fee} as the total fee summed across all initiated transactions, extracted from the \texttt{fee} field; and \emph{asset diversity} as the average number of distinct tokens involved per transaction initiated by the address. 
All features are standardized using z-scores prior to clustering, with missing or non-finite values imputed as 0.
Next, we clustered the bot-associated addresses in our dataset using HDBSCAN~\cite{campello2013density}, 
as it requires no predefined cluster count and handles noise, making it suitable for unknown behavioral bot diversity in our dataset.
The model yielded four clusters of 33, 11, 12, and 102 addresses, and labeled 42 addresses as noise, achieving a silhouette score of 0.6588.

\subsubsection{On-chain Behavior Characterization}
We began with a cluster-wise profitability analysis, and restricted the analysis to transactions where Wrapped SOL (WSOL) serves as the base asset, to enable fair comparison across clusters.
WSOL is the tokenized representation of native SOL, pegged 1:1, which enables SOL to participate in token-based programs and DEX swaps.
Specifically, for each cluster of bot addresses, we examined the distribution of address-level realized profits in SOL-denominated units, which were derived from per-transaction balance deltas between the  \texttt{pre\_token\_balances} and \texttt{post\_token\_balances} fields in transactions.

To interpret the resulting clusters, we further characterize each cluster using observable on-chain operational behaviors, including:
(i) \textbf{Trading venues.} We extract the program ID(s) invoked by each transaction, from both top-level and inner instructions, and map program IDs to venues using Solscan annotations;
(ii) \textbf{Traded tokens.}
We identify the token accounts referenced in swap-related instructions (e.g., input and output token accounts) and extract their mint addresses to determine the traded tokens; 
(iii) \textbf{Transfer outflows.} We extract recipients from native \texttt{transfer} instructions in the System Program and from \path{transfer}/\path{transferChecked} instructions in the Token Program.

To captures economically meaningful transfer outflows that reflect external execution dependencies, rather than internal fund shuffling or protocol-mandated state transitions, we exclude:
(i) intra-transaction transfers that form a closed value loop (i.e., outflows fully returned within the same transaction),
(ii) transfers to bot-controlled addresses used for self-funding (i.e., identified based on publicly annotated funding relationships at the time of data collection), and
(iii) transfers involving protocol state accounts, such as liquidity pool or bonding-curve accounts labeled by Solscan.

\subsubsection{Validation across Time Windows.} \hl{To assess whether the identified behavioral clusters generalize beyond the baseline window and the originally selected bot services, we projected the bot addresses collected during the validation window onto the baseline cluster space, rather than reclustering them. 
The design of projection enables us to evaluate whether the behavioral clusters learned from the baseline window remains applicable to unseen addresses collected in a later time window and from an additional bot-service source. 
Specifically, for each validation address, we computed the same address-level features described in Section 6.1.1. We then assigned each address to the cluster label of its five nearest neighbors in the baseline clustering using a 5-nearest-neighbor classifier with majority voting. Ties were broken by summing inverse distances separately for each tied label and choosing the label with the largest sum~\cite{dudani1976distance}. 
}
\hl{Next, to assess whether the behavioral characteristics of the identified clusters remain stable between the baseline and validation windows, we first characterized the validation addresses assigned to each cluster using the same behavioral dimensions that characterize the baseline clusters, as described in Section 6.1.2. We then compared the resulting characteristics of validation clusters with those of the corresponding baseline clusters in terms of DEX venues, traded tokens, and major outflow recipients.}

\subsection{Results}\label{sec:rq3-results}
\subsubsection{Cluster-level Execution Characteristics}\label{subsec:rq3.1_result}
\begin{figure}
    \centering
    \includegraphics[width=0.75\linewidth, trim={0 2mm 0 0.5mm},clip]{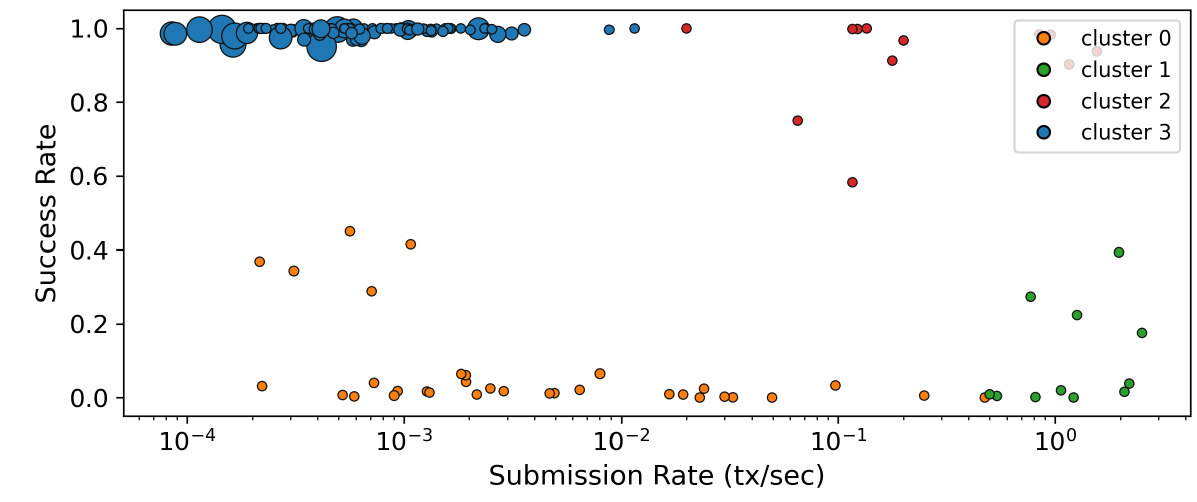}
    \caption{Transaction success rate vs. submission rate across clusters (log-scaled x-axis; bubble size: mean fee per address).}
    \label{fig:tx_rate}
    \vspace{-15pt}
    
\end{figure}

Figure~\ref{fig:tx_rate} visualizes the distribution of bot addresses in the success–frequency space of transaction execution. Rather than forming a single continuous trade-off, the clusters of bot addresses partition the space into four distinct execution regimes, which form a structured 2×2 landscape defined jointly by submission intensity (low vs. high) and execution effectiveness (low vs. high).
At low submission rates (left side of Figure~\ref{fig:tx_rate}), two clusters exhibit sharply different outcomes. \textbf{Cluster~3} concentrates in the upper-left area, achieving near-perfect success \hl{(0.995), with 43.14\% of addresses reaching 100\% success}; while submitting transactions rarely ($9.82\times10^{-4}$ tx/sec), reflecting highly selective execution. In contrast, \textbf{Cluster~0} also operates at very low submission rates \hl{(0.032 tx/sec)} but remains concentrated in the lower-left region, \hl{with 84.85\% of addresses below a 0.2 success rate}, indicating sparse yet ineffective attempts.
At high submission rates (right side of Figure~\ref{fig:tx_rate}), the other two clusters again separate by execution effectiveness. \textbf{Cluster~1} occupies the lower-right region, characterized by frequent submissions \hl{(1.360 tx/sec)} coupled with \hl{72.73\% of addresses below a 0.2 success rate}, consistent with volume-driven speculative behavior. \textbf{Cluster~2}, however, lies in the upper-mid region, combining sustained activity \hl{(0.457 tx/sec)} with high success \hl{(0.918 on average)}, representing a more balanced execution pattern. Table~\ref{tab:tx_rate_summary} in the Appendix provides quantitative summaries of each cluster of the bot addresses.

\begin{figure}[t]
    \centering
    \includegraphics[width=0.8\linewidth]{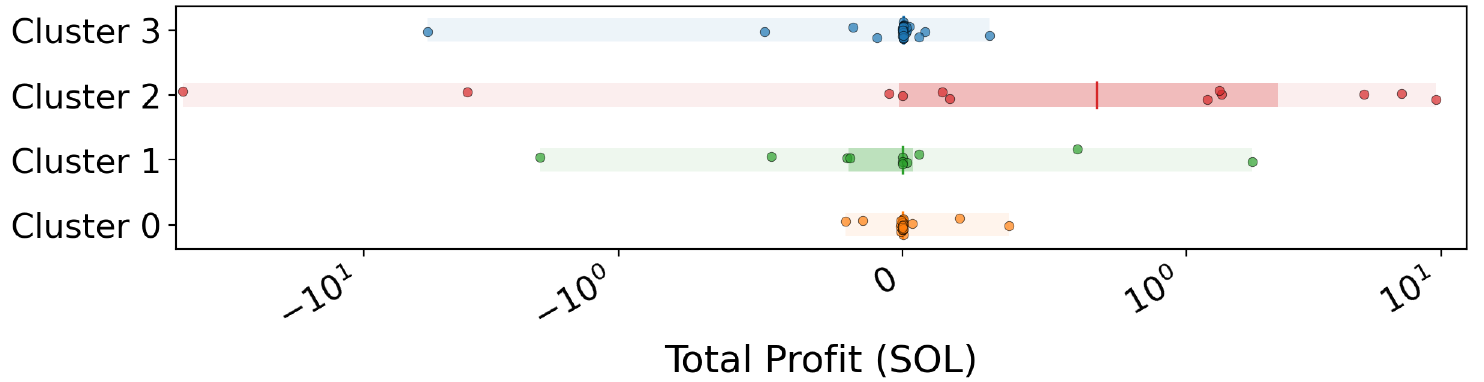}
    \caption{Total profits Distribution across clusters. Bands show the full (light) and interquartile (dark) ranges.}
    \vspace{-17pt}
    \label{fig:Profit}
\end{figure}

We further examined the realized per-address profits to characterize the economic outcomes associated with each execution regime, as shown in Figure~\ref{fig:Profit}.
We observe that execution intensity and success rate do not reliably translate into stable profitability. Instead, they correspond to distinct risk exposures: near-zero outcomes (\textbf{Cluster 0}), frequent small losses (\textbf{Cluster 1}), heavy-tailed downside exposure (\textbf{Cluster 2}), and near-zero gains with occasional downside tail risk (\textbf{Cluster 3}). 
\vspace{-5pt}
\rqbox{
\textbf{Finding 5:} 
Execution characteristics of on-chain bot addresses in the success–frequency space fall into four clusters, characterized by success rate  and submission rate. Realized per-address profit distributions show that neither  execution intensity nor success rate translates into stable profitability, and instead they correspond to distinct risk exposures.
}\vspace{-10pt}

\subsubsection{Trading Venues}\label{sec:trading-venue}

\hl{To interpret the intent-level semantics of the execution regimes identified in Figure~3, we performed transaction-level inspection of representative addresses sampled from each cluster. We operationalized two execution patterns, \emph{round-trip motif} and \emph{venue-specific directional flow}, and manually annotated sampled addresses according to the protocol described below.}
\noindent\hl{\textit{\textbf{Operational Definitions.}} Following prior studies~\cite{wang2022cyclic,mclaughlin2023large} on blockchain arbitrage, we define a \emph{round-trip motif} as a transaction pattern satisfying two conditions: (i) the input and output assets are the same base asset, such as WSOL; (ii) the execution path traverses at least two distinct DEX venues. 
In contrast, we define a \emph{venue-specific directional flow} as a transaction pattern confined primarily to a single venue or a DEX aggregator without a closed-loop return to the base asset, which is consistent with trading-operation behavior.}
\noindent\hl{\textit{\textbf{Sampling and Annotation.}} For each of Clusters~0-3, we stratified addresses into high-, mid-, and low-probability tiers based on HDBSCAN membership probability using a tertile split within each cluster. We randomly sampled up to five addresses per tier; for tiers with fewer than five addresses, all available addresses were included. 
Two annotators independently labeled each address as round-trip, venue-specific directional, or ambiguous/other, achieving 96.23\% inter-annotator agreement. See Appendix~\ref{sec:cluster-mapping} for details.
}

\hl{Across Clusters 0-2, 37 of the 38 sampled addresses exhibited the round-trip motif. One representative transaction~\cite{tx_hash_4MV} executes an aggregated swap across two DEX venues: it first swaps 3.549 WSOL for 23,899.845 PAYAI on Raydium, and then swaps the same amount of PAYAI back for 3.561 WSOL on Meteora DLMM. The resulting WSOL--PAYAI--WSOL path closes on the same base asset while crossing two venues, providing transaction-level evidence of the round-trip arbitrage motif. Meanwhile, all 15 sampled addresses in Cluster 3 exhibited venue-specific directional flows dominated by Pump.fun-related activity. Accordingly, we align Clusters 0–2 with MEV and Cluster 3 with Trading Operations when integrating the clusters into the taxonomy of RQ1.}

\hl{We further cross-validated the semantic interpretation using known bot-service labels. 
Excluding noise, 90.32\% of clustered SolanaMevBot addresses fall in Clusters~0-2 (33, 11, and 12 addresses; 38 noise), whereas all clustered Trojan addresses fall in Cluster~3 (96 addresses; 4 noise). 
This service-level separation corroborates the transaction-level evidence for interpreting Clusters~0--2 as MEV-oriented and Cluster~3 as trading-operation-oriented. 
}

We examined the top 10 trading venues by transaction count across clusters of bot addresses
(Table~\ref{tab:top_dex} in the Appendix).
The four clusters are separated along (i) venue concentration and (ii) aggregator mediation. 
The venue-level differences across clusters also tend to align with the distinct execution regimes observed in Section~\ref{subsec:rq3.1_result}.
Specifically, \textbf{Cluster~2} (Aggregator-Centric MEV), dominated by Jupiter Aggregator v6 (67.9\%), corresponds to the high-success, sustained-activity regime, suggesting that aggregator-mediated routing could be associated with stable execution effectiveness.
In contrast, \textbf{Cluster~3} (Pump.fun-Centric Trading), with over 80\% of transactions confined to the Pump.fun ecosystem, coincides with the low-frequency but near-perfect execution regime,  indicating highly selective activity in a single venue.
\textbf{Cluster~1} (Venue-Specialized MEV), whose activity is concentrated on Raydium and Meteora pools (over 95\% combined), aligns with the high-submission yet low-success regime, reflecting intensive but competitive execution in specialized venues.
\textbf{Cluster~0} (Multi-Venue MEV), which distributes transactions across several venues without strong aggregator reliance, corresponds to the low-activity and low-success regime, indicating dispersed yet ineffective execution.

Beyond venue concentration and aggregator mediation, additional venue-level signals further differentiate the clusters of MEV bot addresses.
In particular, the appearance of Kamino Lending within \textbf{Cluster~0} indicates that lending primitives are occasionally invoked alongside swap instructions, potentially enabling capital-efficient strategies such as flash-loan-assisted arbitrage.
Moreover, \textbf{Cluster~2} interacts with several proprietary AMMs (e.g., HumidiFi~\cite{incrypted_dark_pools_defi,helius_prop_amm_revolution,blockworks_prop_amms}), suggesting routing paths that extend beyond standard public liquidity pools and may involve access to closed or specialized liquidity sources. 
We also observe that, within \textbf{Cluster~2}, transactions invoking the HumidiFi program are associated with more favorable profit outcomes. Specifically, 62.30\% (152,502/244,733) of HumidiFi-invoking transactions are profitable, compared to only 21.01\% (45,949/218,678) of non-HumidiFi-invoking transactions. In addition, HumidiFi-invoking transactions yield an average profit of $143,661$ lamports, whereas non-HumidiFi-invoking transactions incur an average loss of $8,693$ lamports.

\vspace{-8pt}
\rqbox{
\textbf{Finding 6:} 
Trading venues differs across MEV bots. Trading Operations bots confined over 80\% of transactions to the Pump.fun ecosystem.
}\vspace{-12pt}

\subsubsection{Traded Tokens and Transfer Outflows}

We examined the top-10 traded tokens in transactions initiated by each cluster and observed distinct token trading patterns across clusters (Table~\ref{tab:top_tokens} in the Appendix). The main observations are as follows:

\noindent\faRobot~\textbf{\textit{MEV} bots (Clusters 0--2).} WSOL serves as a dominant pivot asset across the MEV-oriented clusters.
WSOL appears in 99.94\% of transactions initiated by addresses in Cluster~1 (Venue-Specialized MEV) and 98.25\% in Cluster~2 (Aggregator-Centric MEV), and remains prevalent in Cluster~0 (Multi-Venue MEV) at 70.90\%.
The clusters also differ in token concentration and breadth.
Cluster~1 exhibits the tightest co-anchor structure, where WSOL (99.94\%) frequently co-occurs with a small set of recurring tokens such as USD1 (46.84\%) and USELESS COIN (44.00\%).
Cluster~0 combines WSOL (70.90\%) with substantial stablecoin exposure (USDC 50.97\%) alongside additional long-tail tokens.
Cluster~2 maintains a WSOL backbone (98.25\%) while spanning a broader mid-frequency token mix (4.13--15.04\% across tokens), indicating wider asset coverage.

\noindent\faRobot~\textbf{\textit{Trading Operations} Bots (Cluster~3)}  show no dominant pivot tokens in their transactions, with WSOL appearing in only 10.93\% of the transactions.
Its top tokens are instead dominated by Pump.fun assets with low individual frequencies (all below 8\%), consistent with a long-tail, platform-driven token distribution in transactions.
Moreover, transfer-outflow recipients suggest different execution-side dependencies across clusters: \textit{MEV} bots are infrastructure-centered, whereas \textit{Trading Operations} bots are more platform- and service-distributed.
Specifically, \textit{MEV} bots (\textbf{Clusters 0--2}) exhibit extreme concentration toward inclusion infrastructure (i.e., Jito tip payment accounts), whereas \textit{Trading Operations} bots (\textbf{Cluster 3}) distribute outflows across trading platforms (e.g., Trojan accounts) and inclusion services (e.g., Jito tip payment accounts). Detailed recipient distributions are provided in Table~\ref{tab:top_recipients_by_cluster} in the Appendix.
\vspace{-5pt}
\rqbox{
\textbf{Finding 7:} 
MEV bots mainly trade WSOL as pivot token, and transfer to infrastructure-related recipients. Trading Operations bots have no dominant pivot, and distribute transfers across platforms and services.}
\vspace{-10pt}

\subsubsection{Cluster and Behavioral Consistency across Time Windows} \hl{The baseline clusters remain broadly applicable to validation-window addresses. All 100 Axiom addresses were mapped to Cluster 3. For 100 SolanaMevBot addresses from the validation window, 41, 4, and 2 were assigned to Clusters 0, 1, and 2, respectively; 20 were assigned to Cluster 3; and the remaining 33 were classified as noise. This corresponds to 0\% and 33\% noise ratios for Axiom and SolanaMevBot addresses in the validation window, respectively. The remaining SolanaMevBot noise addresses suggest that some later-window behaviors are less well captured by the clusters observed in the baseline window. Details see Appendix~\ref{sec:validation-window}.
}

\hl{Different behavioral dimensions exhibit different levels of temporal stability across the baseline and validation windows. Overall, trading-venue usage remains the most stable behavioral dimension, followed by transfer-outflow recipients, whereas traded tokens exhibit substantially greater temporal variation.}
\hl{ \textit{\textbf{Trading Venues.}} For the MEV-oriented clusters, C0 preserves its original multi-venue pattern, with 7 of its top-10 DEX venues overlapping with the original C0, including Jupiter, Pump.fun AMM, Meteora DLMM, and Whirlpools. C1 exhibits the strongest continuity, with 8 overlapping top-10 venues centered on Meteora, Raydium, Pump.fun AMM, and Whirlpools. C3 also retains a Pump.fun-oriented venue profile, with Pump.fun and Pump.fun AMM remaining among the dominant venues. Consistently, all validation-window Axiom addresses, which are assigned to C3, exhibit the same Pump.fun-oriented trading behavior. In contrast, C2 shows weaker stability, with only 2 overlapping top-10 venues.}
\hl{\textit{\textbf{Transfer-outflow Recipients.}} C1 and C2 remain strongly Jitotip-centered, accounting for 80.64\% and 100\% of total transfers, respectively. C3 continues to direct transfers primarily to Jitotip (18.76\%) and Pump.fun protocol-fee accounts (14.39\%). Although C0 shows a different set of dominant recipient accounts, it still shares common service-level recipients such as Pump.fun protocol-fee accounts, indicating that the underlying execution infrastructure remains largely unchanged.}
\hl{\textit{\textbf{Traded Tokens}} exhibit substantially greater temporal variation. Across clusters, only 3 of the top-10 traded tokens overlap with those in the baseline window. The stable overlap mainly consists of common base or pivot assets, including WSOL, USDC, and USD1, whereas long-tail traded tokens change considerably across the validation window.
}
\vspace{-15pt}
\section{Implications}\label{sec:discussion}
\noindent\textbf{Architectural Regularity and Engineering Implications for Solana Bots}.
\hl{Solana bots exhibit a stable high-level execution backbone despite substantial variation in low-level implementation (Finding 2)}. 
\hl{At the architectural level, repositories across bot categories repeatedly align with a common five-stage pipeline, especially around shared components such as \textit{Data Acquisition}, \textit{Data Analytics}, and \textit{Trade Execution} (Finding 3)}.
The observed architectural regularity of Solana bots at the code level is broadly consistent with prior studies of bots on Ethereum and BNB~\cite{cernera2023token,niedermayer2024detecting}, which identify recurring operational motifs such as sniping, front-running, and arbitrage from account- or transaction-level evidence.
Meanwhile, the recurrence of a small set of building blocks suggests that reimplementation is pervasive in the Solana bot OSS ecosystem. 

\noindent \hl{\takeaway{Instead of repeatedly developing end-to-end bots from scratch, practitioners should prioritize the shared maintenance of recurrent bot components, while preserving flexibility for long-tail, strategy-specific logic. Moreover, the five-stage pipeline identified in RQ2 can be viewed as an empirically derived architectural reference model for designing and auditing Solana bots.}}

\noindent\textbf{Dependency Lag as an Ecosystem-Level Maintainability Concern of Solana Bots}.
\hl{More than 30\% of the dependencies in the third-party libraries in Solana lag behind the latest available release by over one year (Finding 4). }
The observation suggests that dependency technical lag is not merely a repository-level issue, but an ecosystem-level maintainability concern for Solana bots.  
\hl{The maintainability concern, however, is not uniform across dependency types: for protocol-facing libraries such as core Solana SDKs, lag poses a genuine compatibility risk, since Solana's transaction formats, RPC interfaces, and fee-setting mechanisms evolve continuously, and staleness here risks silent incompatibility rather than merely missing new features. For general-purpose libraries not directly coupled to protocol evolution (e.g., dotenv, axios), lag instead reflects a stability-versus-currency tradeoff familiar from software maintenance more broadly, where a pinned, well-tested dependency environment can be preferable to frequent, disruptive upgrades.}
The maintainability concern is especially salient for Solana SDK dependencies, \hl{where the technical lag of widely used Solana SDKs ranges from 125 to 572 days. }
Blockchain bots tend to operate in latency-sensitive and adversarial transaction environments, where their execution success depends heavily on SDK APIs, RPC interfaces, transaction formats, and fee-setting mechanisms~\cite{mclaughlin2023arbitrage,gerzon2025jito}. 

\noindent \hl{\takeaway{Practitioners, therefore, should prioritize the management of SDK dependencies in Solana bots as part of ensuring transaction correctness and execution reliability, rather than as a background maintenance task, while general-purpose dependencies can be managed with more discretion, favoring stability where frequent upgrades offer limited benefit. Correspondingly, tool builders should support not only outdated-dependency detection, but also compatibility assessment for evolving transaction formats, RPC interfaces, and fee-related mechanisms, so that update decisions can be prioritized by dependency type rather than applied uniformly.}}
Future work could investigate the underlying drivers of dependency lag in Solana bot repositories, including how the evolution of SDK libraries and transaction mechanisms affects dependency update decisions, maintenance effort, and compatibility management.

\noindent\textbf{\hl{The Interaction of Infrastructure, Venue, and Execution Regime is Associated with Distinct Execution Outcomes of Solana Bots}}.
Solana bots can be characterized through three interacting layers: (1) execution infrastructure, such as Jito, which shapes transaction submission and ordering; (2) DEX venue integration, such as proprietary AMM (e.g., HumidiFi), which \hl{is associated with } how bots access liquidity and routing opportunities; and (3) execution regimes, which reflect how bots realize trading strategies within these constraints. Execution outcomes, therefore, are \hl{affected by} the interaction of execution infrastructure, venue integration, and execution regime, rather than by trading logic alone (Findings 6 and 7), which is consistent with prior studies~\cite{mclaughlin2023arbitrage,gerzon2025jito}.

\noindent\takeaway{Practitioners should design Solana bots through joint consideration of execution infrastructure, venue integration, and execution regimes, to secure execution edges in the highly competitive Solana market. Tool builders could support the design exploration across these layers.}
Future work could investigate the mechanisms underlying the observed coupling among execution infrastructure, venue integration, and execution regimes, including whether this coupling reflects deliberate design choices by practitioners, latency-driven optimization pressures~\cite{daian2020flashboys, gerzon2025jito}, or access constraints imposed by private liquidity ecosystems~\cite{helius_prop_amm_revolution}. 
Cross-chain comparisons could further examine how ordering mechanisms, mempool visibility, and fee structures shape bot specialization.

\vspace{-10pt}
\section{Threats to Validity}
\noindent\hl{\textbf{Construct Validity.} Our GitHub repository and on-chain address datasets of Solana bots were collected independently and cannot be linked at the address level.
The two datasets therefore provide complementary views of the Solana bot ecosystem, covering implementation-level functionality and execution-level behavior.
Accordingly, in Section~\ref{sec:trading-venue}, we establish correspondence between the two datasets at the behavioral level by linking on-chain clusters to the MEV and Trading Operations categories based on transaction traces from representative addresses in each cluster.
}

\noindent\hl{\textbf{External Validity.} Three repository domains (\textit{Market Manipulation}, \textit{On-chain Analytics}, and \textit{Tooling and Infrastructure}) are absent from our on-chain cluster, because the two analyzed bot services (Trojan and SolanaMevBot) are trading/arbitrage-oriented and do not exercise these domains. 
The observation reflects the service-level scope of our on-chain sample rather than an artifact of the clustering procedure. Consequently, our on-chain findings (Section~\ref{sec:rq3-results}) generalize primarily to trading- and MEV-oriented bots.
}

\vspace{-10pt}
\section{Related Work}\label{sec:related}
\noindent\textbf{Studies on Solana.}
Early work investigated the performance and scalability of the Solana blockchain through its architecture~\cite{li2021bitcoin}, capability for large-scale IoT workloads~\cite{duffy2021IoT}, and  transaction latency and 
fees~\cite{pierro2022scalability}.
Other prior work developed automated techniques to detect vulnerabilities in Solana smart contracts using static analysis~\cite{vrust} and coverage-guided fuzzing~\cite{smolka2023fuzz}.
Recent studies have examined diverse on-chain phenomena on Solana through systematic measurement, including failed transactions~\cite{zheng2025does}, MEV sandwiching attacks~\cite{gerzon2025quantifying}, meme coin markets~\cite{li2025trust}, and rug pulls~\cite{alhaidari2025solrpds}, providing insights into transaction-level behaviors and economic dynamics on Solana.
Despite extensive prior work on Solana, existing studies overlook Solana bots as software artifacts despite generating substantial on-chain activity.
We addresses this gap through a systematic empirical study of Solana bots \hl{that jointly examines their code-level characteristics and observable on-chain behaviors.}

\noindent\textbf{Studies on Blockchain Bots.}
One line of prior work examines the behaviors of bots in AMM ecosystems on the Ethereum and BSC blockchains, including token sniping~\cite{cernera2023sniper, cernera2025blockchain}, front-running~\cite{daian2020flash, qin2022quantifying, li2023towards}, and sandwich attacks~\cite{qin2022quantifying, zhou2021high, torres2021frontrunner}. 
Another line of prior work explores the automatic detection of bot-controlled accounts across blockchain platforms, such as EOSIO~\cite{huang2020eosio} and Ethereum~\cite{niedermayer2024bots, jin2022detecting}. 
A recent study characterizes the bot accounts on the Solana blockchain that initiate failed transactions~\cite{zheng2025does}. 
Overall, existing studies primarily focus on account-level analyses of bot behaviors, with most efforts centered on the Ethereum and BSC blockchains.
In contrast, our work explores bots on the Solana blockchain from a software-centric perspective.
Closely related to our work, Niedermayer et al.~\cite{niedermayer2024bots} derive an Ethereum financial-bot taxonomy from literature, GitHub repositories, and on-chain data. The taxonomy organizes financial bot activity into 7 categories and 24 subcategories, which elevates \textit{MEV} as a mechanism-level category rooted in transaction-ordering competition, distinguishes venue-based branches (i.e., \textit{CEX} and \textit{DEX}), and further differentiates application- or asset-oriented domains such as \textit{NFT} and \textit{Play-to-earn}.

\vspace{-5pt}
\section{Conclusion and Future Work}\label{sec:conclusion}


In this work, we present the first large-scale, implementation-grounded characterization of Solana bots by jointly analyzing 586 GitHub repositories and 200 bot-associated addresses with over 44 million on-chain transactions. 
We derived a 15-category taxonomy, identified a shared five-stage operational
pipeline of bot implementations, and found distinct bot execution fingerprints
across venues and assets.
Our findings \hl{characterize Solana bots from both code-level implementation and observable on-chain execution perspectives}, and highlight relevant software engineering concerns that arise in architecture, dependency management, and cross-layer execution design for bots operating in a competitive execution environment and a fast-evolving software stack.
Future work could explore the differences in ordering mechanisms, mempool visibility, and fee structures across blockchain systems, and how they shape distinct forms of bot specialization across blockchain systems. 

\vspace{-5pt}
\section{Data Availability}\label{sec:repo}
Our replication package is available online:
\hl{\url{https://doi.org/10.5281/zenodo.21359451}}.

\appendix
\appendix
\onecolumn
\section{Appendix}\label{sec:appendix}
\subsection{Representative Solana Bot Repositories}
\faRobot~The \texttt{nmweaver/soltrade} repository, written in Python, exemplifies the \textit{Order Execution and Management} category by implementing a closed-loop execution logic for a Solana bot: the bot periodically monitors market data, derives trading signals using technical indicators, and plans for executing routed swaps via Jupiter.\footnote{Jupiter is a widely used DEX aggregator on Solana that computes optimal swap routes across multiple DEXs and provides ready-to-submit swap transactions.} At runtime, once a signal is triggered, the bot obtains a swap route and a serialized transaction from Jupiter, then submits the transaction with a fixed priority fee (via compute-unit pricing) to improve the probability of on-chain inclusion.

\noindent\faRobot~The \texttt{warp-id/solana-trading-bot} repository, written in TypeScript, exemplifies the \emph{Sniping} category by implementing an event-driven strategy that monitors newly created Raydium liquidity pools,\footnote{Raydium is a DEX and automated market maker (AMM) protocol on Solana, where liquidity pools facilitate token swaps.} and promptly submits \emph{buy} transactions for eligible pools. The bot restricts sniping to pools quoted in configured assets (e.g., USDC or WSOL), and performs pre-trade checks (e.g., skipping pools created before startup and, when enabled, only trading pools on an allowlist) before submitting a \textit{buy} transaction.
The bot exposes multiple transaction submission backends (e.g., Warp and Jito), reflecting the use of alternative low-latency transaction relay or accelerator services to achieve timely on-chain inclusion of transactions.

\begin{table*}[ht]
\centering
\scriptsize
\caption{Taxonomy of Solana Bot Repositories (Repos) on GitHub.}
\label{tab:taxonomy_detailed}
\rowcolors{2}{gray!10}{white}
\renewcommand{\arraystretch}{1.2}
\begin{tabular}{p{2.2cm} p{8.6cm} p{3.2cm} p{2.2cm}}
\toprule
\textbf{Category (\# Repos)} & \textbf{Description} & \textbf{Example Tags}
& \textbf{Representative Repo} \\
\hline

\multicolumn{4}{l}{\cellcolor{lightgray}\textbf{\faRobot~Trading Operations}}\\
\textbf{Order Execution and Management (137)}
&
Bots that automate on-chain order placement with execution control~\cite{fang2022cryptocurrency}, enabling conditional or triggered orders, scheduled or split execution, rule-based repeated trading, and safeguards such as slippage limits or minimum-output constraints.
&
\textit{trading-bot}, \textit{automated-trading}, \textit{transaction-execution}, \textit{limit-orders}, \textit{telegram-trading-bot}
&
\path{nmweaver/soltrade} (342~\tiny \faStar)
\\

\textbf{Liquidity Provision and Yield Farming (12)}
&
Bots that automate liquidity position management on DEX by providing or removing liquidity~\cite{milionis2022automated}, periodically rebalancing positions, harvesting rewards, and automatically reinvesting them, and optionally applying safeguards to reduce exposure to impermanent loss.
&
\textit{liquidity-management}, \textit{bonding-curve}, \textit{liquidity-bot}, \textit{automated-market-maker}, \textit{liquidity-pools}
&
\path{edwin-finance/meteora-liquidity-rebalancer} (4~\tiny \faStar) 
\\

\textbf{Copy Trading (33)}
&
Bots that automate trade mirroring by monitoring designated ``leader'' wallets and replicating their positions or transactions~\cite{kawai2024stranger,apesteguia2020copy}, often framed as social/smart-money-following trading.
&
\textit{copy-trading}, \textit{copy-trading-bot}
&
\path{cryptole0/Copy-Trading-Bot-Rust} (96~\tiny \faStar)
\\

\hline
\multicolumn{4}{l}{\cellcolor{lightgray}\faRobot~\textbf{MEV}}\\
\textbf{Arbitrage (47)}
&
Bots that automate profit-seeking trades by exploiting price or rate discrepancies across trading venues (e.g., cross-DEX, CEX--DEX, and flash-loan amplified paths), via fast, often multi-leg execution~\cite{mclaughlin2023arbitrage,wang2022cyclic}.
&
\textit{arbitrage-bot}, \textit{flash-loan-arbitrage}, \textit{cross-pool-arbitrage}, \textit{cross-exchange-arbitrage}, \textit{dex-cex-arbitrage}
&
\path{0xNineteen/solana-arbitrage-bot} (799~\tiny \faStar)
\\

\textbf{Sniping (134)}
&
Bots that automate event-driven trading around new token launches by monitoring on-chain signals and launch platforms~\cite{cernera2023token,cernera2023ready}, which rapidly submit buy/sell transactions during the initial liquidity and early trading window to secure early entry.
&
\textit{sniper-bot}, \textit{pumpfun-bot}, \textit{pumpfun-trading-bot}, \textit{memecoin-trading}, \textit{token-sniper}
&
\path{warp-id/solana-trading-bot} (2,220~\tiny \faStar)

\\

\textbf{Transaction Ordering Exploitation (41)}
&
Bots that automate ordering-based execution by shaping in-block/slot transaction inclusion and ordering~\cite{yang2024sokmev}, using bundle construction and simulation, and bidding for priority (e.g., via Jito tips) to enable ordering-dependent strategies such as front-running and sandwiching.
&
\textit{mev-bot}, \textit{jito-bundle}, \textit{transaction-bundling}, \textit{pumpfun-bundler}, \textit{sandwich-attack}
&
\path{jito-labs/mev-bot } (1,150~\tiny \faStar)
\\

\hline
\multicolumn{4}{l}{\cellcolor{lightgray}\textbf{\faRobot~Market Manipulation}}\\
\textbf{Wash Trading (21)}
&
Bots that automate the generation of artificial trading activity~\cite{volumebot,volumebot2} by creating high-frequency buy/sell flows that inflate apparent volume, mimic organic trading patterns (e.g., randomized timing and order sizes), and spread activity across multiple transactions or addresses to reduce detectability.
&
\textit{volume-bot}, \textit{volume-simulation}
&
\path{web3batman/Raydium-Volume-Bot} (67~{\tiny\faStar})
\\

\hline
\multicolumn{4}{l}{\cellcolor{lightgray}\textbf{\faRobot~On-chain Analytics}}\\
\textbf{Monitoring and Alerting (26)}
&
Bots that monitor and analyze on-chain activity in real time by subscribing to blockchain events (e.g., liquidity changes, wallet flows, and new token listings), and publish alerts to support downstream decision-making or trigger automated trading~\cite{solana-aml-bot,solana-aml-bot2}.
&
\textit{real-time-alerts}, \textit{token-monitoring}, \textit{real-time-monitoring}, \textit{wallet-tracking}, \textit{transaction-monitoring}
&
\path{FriedDev/solana-rug-checker} (29~{\tiny\faStar})
\\

\textbf{Portfolio and Risk Management (14)}
&
Bots that automate portfolio and risk management by tracking holdings and performance (e.g., profit and loss), rebalancing portfolios, and evaluating strategies through backtesting and simulation~\cite{urusov2025backtesting,zuniga2023maximizing}.
&
\textit{risk-management}, \textit{risk-analysis}, \textit{portfolio-management}, \textit{simulation}, \textit{data-analysis}
&
\path{henrytirla/Solana-PNL-Bot} (66~{\tiny\faStar})
\\

\hline
\multicolumn{4}{l}{\cellcolor{lightgray}\textbf{\faCogs~Tooling and Infrastructure}}\\
\textbf{Token Issuance and Administration (8)}
&
Tools that automate token issuance and administration~\cite{token-automation} (SPL/Token-2022), including creation/deployment, minting and distribution, metadata setup/validation, and administrative controls.
&
\textit{token-launchpad}, \textit{token-creation}, \textit{token-launch}, \textit{token-management}, \textit{spl-token}
&
\path{0xNevo/Solana_Token_Freezer} (14~{\tiny\faStar})

\\

\textbf{NFT Minting and Marketplace Trading (14)}
&
Tools that automate NFT issuance and marketplace operations~\cite{nft-minting,nft-minting2} by minting items, monitoring listings and sales activity on marketplaces, and generating transactions for listing, buying, or selling.
&
\textit{nft-sales-bot}, \textit{nft-tools}, \textit{nft-minting-bot}, \textit{nft}, \textit{nft-bot}
&
\path{theskeletoncrew/air-support} (178~{\tiny\faStar})
\\

\textbf{DEX Swap Routing and Integration (30)}
&
Tools provide reusable components and services for token swapping across trading venues (AMMs, order books, and aggregators), including API integrations, multi-hop routing, liquidity aggregation, and optional cross-chain bridging support~\cite{zhang2025line,jupiter-routing}.
&
\textit{raydium-integration}, \textit{jupiter-aggregator}, \textit{dex-integration}, \textit{jupiter-swap}, \textit{dex-aggregator}
&
\path{YZYLAB/solana-swap} (130~{\tiny\faStar})
\\

\textbf{Wallet Management (6)}
&
Tools automate wallet and account operations such as keypair and token-account management, multi-wallet coordination, transaction building and batching, and fund distribution/sweeping~\cite{bulk-transfer,multi-wallets}.
&
\textit{wallet-management}, \textit{multi-wallet}, \textit{token-transfer}, \textit{account-management}, \textit{keypair-management}
&
\path{LeaderMalang/Solana-Sweeper-Bot} (9~{\tiny\faStar})
\\

\textbf{On-chain Data Pipeline (10)}
&
Infrastructure and tools for collecting, streaming, and indexing on-chain data via RPC, Geyser, gRPC, or webhooks for downstream queries and applications.
&
\textit{rpc-client}, \textit{data-aggregation}, \textit{birdeye-api}, \textit{geyser-client}, \textit{helius-webhook}
&
\path{weeaa/goyser} (66~{\tiny\faStar})
\\

\textbf{Messaging Integration and Alerts (37)}
&
Tools integrate bot capabilities with chat and social platforms (e.g., Telegram, Discord, and Twitter), providing event-driven alerts via webhooks and interactive command interfaces for downstream automation and coordination.
&
\textit{telegram-bot}, \textit{twitter-integration}, \textit{discord-bot}, \textit{chat-interface}, \textit{discord-notifications}
&
\path{KingJiongEN/DegentGroup} (65~{\tiny\faStar})
\\

\bottomrule
\end{tabular}
\end{table*}

\begin{table}[t]
\centering
\scriptsize
\caption{\hl{Repository-level dependency lag across bot categories.}}
\label{tab:category_dependency_lag}
\renewcommand{\arraystretch}{0.9}
\setlength{\tabcolsep}{2.2pt}
\resizebox{0.45\linewidth}{!}{
\begin{tabular}{@{}p{1.45cm}p{3.25cm}rr@{}}
\toprule
\textbf{Domain} & \textbf{Category} & \textbf{\makecell{Median\\(days)}} & \textbf{\makecell{IQR\\(days)}} \\
\midrule
\multirow{3}{1.45cm}{\makecell[l]{Trading\\Operations}}
  & Order Execution and Management              & 4.0   & 125.8 \\
  & Liquidity Provision and Yield Farming       & 121.3 & 251.3 \\
  & Copy Trading                                 & 0.0   & 65.9  \\
\midrule
\multirow{3}{1.45cm}{MEV}
  & Arbitrage                                    & 391.2 & 435.6 \\
  & Sniping                                      & 0.0   & 125.8 \\
  & Transaction Ordering Exploitation            & 0.0   & 159.1 \\
\midrule
\makecell[l]{Market\\Manipulation}
  & Wash Trading                                 & 62.9  & 125.8 \\
\midrule
\multirow{2}{1.45cm}{\makecell[l]{On-chain\\Analytics}}
  & Monitoring and Alerting                      & 0.0   & 125.8 \\
  & Portfolio and Risk Management                & 202.2 & 679.9 \\
\midrule
\multirow{6}{1.45cm}{\makecell[l]{Tooling and\\Infrastructure}}
  & Token Issuance and Administration            & 0.2   & 265.7 \\
  & NFT Minting and Marketplace Trading          & 433.9 & 629.6 \\
  & DEX Swap Routing and Integration             & 0.0   & 91.9  \\
  & Wallet Management                            & 151.3 & 267.6 \\
  & On-chain Data Pipeline                       & 0.0   & 136.3 \\
  & Messaging Integration and Alerts             & 62.9  & 238.5 \\
\bottomrule
\end{tabular}
}
\vspace{-8pt}
\end{table}

\subsection{Validation Dataset of On-chain Addresses of Solana Bots}\label{sec:validation-window}
The transaction volume per address during the validation window also varies substantially across the two bot services. Bot addresses from SolanaMevBot range from 1 to 6,999,255 transactions (median: 1,651.5; mean: 464,077), whereas bot addresses from Axiom range from 454 to 40,270 transactions (median: 5,187; mean: 10,200).

The noise-like SolanaMevBot addresses showed mixed boundary profiles. They partially resembled baseline Cluster 1 in exhibiting high-frequency, low-success behavior, with comparable transaction frequency (1.78 vs. 1.32 transactions per block). However, they were less extreme than the Cluster 1, with lower transaction volume (median 121,356 vs. 1,658,441 transactions) and a higher, though still low, success rate (6.91\% vs. 1.95\%). 

\subsection{Validation of  Cluster-to-Intent Mapping} \label{sec:cluster-mapping}
The stratified sampling method covers both highly prototypical addresses and boundary cases near the periphery of each cluster. 
Each sampled address was independently labeled by two annotators as exhibiting a round-trip motif, a venue-specific directional flow, or ambiguous/other, according to the operational definitions in Section~\ref{sec:trading-venue}. Inter-annotator agreement reached 96.23\% by percentage agreement. Addresses that matched neither pattern were labeled as ambiguous/other and reported separately.
Table~\ref{tab:manual-inspection} summarizes the manual validation results for the intent-level semantics of Clusters 0-3.
\begin{table}[t]
\centering
\scriptsize
\renewcommand{\arraystretch}{0.8}
\setlength{\tabcolsep}{4pt}
\caption{Quantitative transaction summaries by cluster. Fees are in lamports.}
\label{tab:tx_rate_summary}
\resizebox{0.55\linewidth}{!}{%
\begin{tabular}{llrrrr}
\toprule
\textbf{Cluster} & \textbf{Metric} & \textbf{Min} & \textbf{Median} & \textbf{Mean} & \textbf{Max} \\
\midrule
\multirow{3}{*}{0} & Success Rate & 0.000 & 0.017\% & 0.073 & 0.450 \\
& Transaction Fee     & 5,039 & 7,636 & 17,692 & 99,211 \\
& Submission Rate   & 0.000 & 0.002 & 0.032 & 0.475 \\
\midrule
\multirow{3}{*}{1} & Success Rate & 0.000 & 0.020 & 0.105 & 0.393 \\
& Transaction Fee     & 5,384 & 5,730 & 7,076 & 14,624 \\
& Submission Rate   & 0.500 & 1.219 & 1.360 & 2.520 \\
\midrule
\multirow{3}{*}{2} & Success Rate & 0.583 & 0.975 & 0.918 & 1.000 \\
& Transaction Fee     & 5,005 & 7,633 & 11,239 & 36,415 \\
& Submission Rate   & 0.020 & 0.157 & 0.457 & 1.561 \\
\midrule
\multirow{3}{*}{3} & Success Rate & 0.950 & 0.999 & 0.995 & 1.000 \\
& Transaction Fee     & 5,000 & 131,078 & 924,123 & 7,302,435 \\
& Submission Rate   & $0.85\times10^{-4}$ & $5.61\times10^{-4}$ & $9.82\times10^{-4}$ & 0.012 \\
\bottomrule
\end{tabular}
}
\end{table}

\begin{table}[t]
\centering
\caption{\hl{Validation of Cluster Intent-Level Semantics.}}
\label{tab:manual-inspection}
\resizebox{0.5\linewidth}{!}{%
\begin{tabular}{crrrr}
\toprule
Cluster & \#Sampled & Round-trip motif & Directional flow & Ambiguous/Other \\
\midrule
C0 & 15 & 14 (93.33\%) & 0 (0.00\%) & 1 (6.67\%) \\
C1 & 11 & 11 (100.00\%) & 0 (0.00 \%) & 0 (0.00 \%) \\
C2 & 12 & 12 (100.00\%) & 0 (0.00\%) & 0 (0.00 \%) \\
C3 & 15 & 0 (0.00 \%) & 15 (100.00\%) & 0 (0.00 \%) \\
\bottomrule
\end{tabular}
}
\end{table}

\begin{table*}[ht]
\centering
\footnotesize
\setlength{\tabcolsep}{2.5pt}
\renewcommand{\arraystretch}{0.9}
\caption{Top-10 trading venues by transaction count across clusters. We denote Cluster~0–3 as C0–3 for brevity.
}
\label{tab:top_dex}
\begin{subtable}[t]{0.24\linewidth}
\centering
\caption{C0: Multi-Venue MEV}
\begin{tabular}{ll}
\toprule
 \textbf{Venue (Program)} & \textbf{Txs \# | \%} \\
\midrule
\cellcolor{lightgray!20}Pump.fun & 2,221 | 33.02\% \\
\cellcolor{lightgray!20}Whirlpools & 2,058 | 30.59\% \\
\cellcolor{lightgray!20}Raydium CLMM & 2,058 | 30.59\% \\ 
\cellcolor{lightgray!40}Jupiter Aggregator v6 & 741 | 11.02\% \\
\cellcolor{lightgray!20}Kamino Lending & 670 | 9.96\% \\
\cellcolor{lightgray!20}Pump.fun AMM & 511 | 7.60\% \\
\cellcolor{lightgray!20}Raydium CPMM & 369 | 5.49\% \\
\cellcolor{lightgray!40}Meteora DLMM & 367 | 5.46\% \\
\cellcolor{lightgray!20}Raydium Liquidity & 151 | 2.24\% \\
\cellcolor{lightgray!20}HumidiFi & 84 | 1.25\% \\
\bottomrule
\end{tabular}
\end{subtable}
\hfill
\begin{subtable}[t]{0.24\linewidth}
\centering
\caption{C1: Venue-Specialized MEV}
\begin{tabular}{ll}
\toprule
\textbf{Venue (Program)} & \textbf{Txs \# | \%} \\
\midrule
\cellcolor{lightgray!40}Meteora DLMM & 1,338,569 | 34.56\% \\
\cellcolor{lightgray!20}Raydium Liquidity & 1,261,944 | 32.58\% \\
\cellcolor{lightgray!20}Raydium CPMM & 1,129,755 | 29.16\% \\
Meteora DAMM v2 & 83,794 | 2.16\% \\
Meteora Pools & 33,559 | 0.87\% \\
\cellcolor{lightgray!20}Pump.fun AMM & 20,710 | 0.53\% \\
\cellcolor{lightgray!20}Whirlpool & 10,513 | 0.27\% \\
\cellcolor{lightgray!20}Raydium CLMM & 7,783 | 0.20\% \\
Heaven DEX & 4,531 | 0.12\% \\
\cellcolor{lightgray!40}Jupiter Aggregator v6 & 4,192 | 0.11\% \\
\bottomrule
\end{tabular}
\end{subtable}
\hfill
\begin{subtable}[t]{0.24\linewidth}
\centering
\caption{C2: Aggregator-Centric MEV}
\begin{tabular}{ll}
\toprule
\textbf{Venue (Program)} & \textbf{Txs \# | \%} \\
\midrule
\cellcolor{lightgray!40}Jupiter Aggregator v6 & 314,641 | 67.90\% \\
\cellcolor{lightgray!20}HumidiFi & 244,733 | 52.81\% \\
\cellcolor{lightgray!20}Whirlpool & 162,116 | 34.98\% \\
\cellcolor{lightgray!40}Meteora DLMM  & 147,015 | 31.72\% \\
SolFi & 69,461 | 14.99\% \\
ZeroFi & 66,539 | 14.36\% \\
Obric V2 & 64,037 | 13.82\% \\
Jupiter Perpetuals & 63,057 | 13.61\% \\
\cellcolor{lightgray!20}Kamino Lending & 62,500 | 13.49\% \\
Tessera V & 56,859 | 12.27\% \\
\bottomrule
\end{tabular}
\end{subtable}
\hfill
\begin{subtable}[t]{0.24\linewidth}
\centering
\caption{C3: Pump.fun-Centric Trading}
\begin{tabular}{ll}
\toprule
\textbf{Venue (Program)} & \textbf{Txs \# | \%} \\
\midrule
\cellcolor{lightgray!20}Pump.fun AMM & 58,517 | 46.26\% \\
\cellcolor{lightgray!20}Pump.fun & 43,812 | 34.63\% \\
Meteora DAMM & 7,765 | 6.14\% \\
\cellcolor{lightgray!40}Jupiter Aggregator v6 & 6,379 | 5.04\% \\
\cellcolor{lightgray!20}Raydium Liquidity & 5,756 | 4.55\% \\
\cellcolor{lightgray!40}Meteora DLMM & 5,011 | 3.96\% \\
\cellcolor{lightgray!20}Raydium CPMM & 4,046 | 3.20\% \\
SolFi V2 & 1,917 | 1.52\% \\
{\scriptsize Meteora Dynamic Bonding Curve} & 1,562 | 1.23\% \\
\cellcolor{lightgray!20}Raydium CLMM & 1,105 | 0.87\% \\

\bottomrule
\end{tabular}
\end{subtable}
\end{table*}

\begin{table*}[t]
\centering
\footnotesize
\setlength{\tabcolsep}{2.5pt}
\renewcommand{\arraystretch}{0.9}
\caption{Top-10 traded tokens by transaction count across clusters.}

\label{tab:top_tokens}
\begin{subtable}[t]{0.24\linewidth}
\centering
\caption{C0: Multi-Venue MEV}
\begin{tabular}{>{\raggedright\arraybackslash}p{0.55\linewidth} >{\raggedleft\arraybackslash}p{0.4\linewidth}}
\toprule
\textbf{Token Mint} & \textbf{Txs \# | \%} \\
\midrule
WSOL & 19,351 | 70.90\% \\
USDC & 13,911 | 50.97\% \\
DraperTV (Pump.fun)& 3,816 | 13.98\% \\ 
Meteora & 2,146 | 7.86\% \\
Lets go (Pump.fun)& 1,815 | 6.65\% \\ 
DoubleZero & 1,318 | 4.83\% \\
USD1 & 1,170 | 4.29\% \\
SLERF & 1,015 | 3.72\% \\
1\textsuperscript{a}  & 838 | 3.07\% \\
@easytopredict  & 715 | 2.62\% \\
\bottomrule
\end{tabular}
\end{subtable}
\hfill
\begin{subtable}[t]{0.24\linewidth}
\centering
\caption{C1: Venue-Specialized MEV}
\begin{tabular}{>{\raggedright\arraybackslash}p{0.5\linewidth} >{\raggedleft\arraybackslash}p{0.45\linewidth}}
\toprule
\textbf{Token Mint} & \textbf{Txs \# | \%} \\
\midrule
WSOL & 23,674,438 | 99.94\% \\
USD1 & 11,096,721 | 46.84\% \\
USELESS COIN & 10,422,887 | 44.00\% \\ 
GMvC...bonk & 7,672,218 | 32.39\% \\ 
\footnotesize{Sombrero Memes} & 6,211,416 | 26.22\% \\ 
Fartcoin (Pump.fun)& 3,727,453 | 15.73\% \\ 
USDC & 3,358,702 | 14.18\% \\
PAYAI (Pump.fun)& 1,675,679 | 7.07\% \\
\$1 Trump Coin & 1,104,440 | 4.66\% \\
@easytopredict & 702,145 | 2.96\% \\
\bottomrule
\end{tabular}
\end{subtable}
\hfill
\begin{subtable}[t]{0.24\linewidth}
\centering
\caption{C2: Aggregator-Centric MEV}
\begin{tabular}{>{\raggedright\arraybackslash}p{0.5\linewidth} >{\raggedleft\arraybackslash}p{0.45\linewidth}}
\toprule
\textbf{Token Mint} & \textbf{Txs \# | \%} \\
\midrule
WSOL & 11,431,418 | 98.25\% \\
USELESS COIN & 1,750,295 | 15.04\% \\
@easytopredict  & 728,599 | 6.26\% \\
\footnotesize{KITKAT (Pump.fun)} & 712,506 | 6.12\% \\
PAYAI (Pump.fun) & 703,558 | 6.05\% \\
USDC & 659,232 | 5.67\% \\
FUN & 568,958 | 4.89\% \\
USD1 & 513,994 | 4.42\% \\
KURO & 501,682 | 4.31\% \\
\footnotesize{NEO} & 479,972 | 4.13\% \\
\bottomrule
\end{tabular}
\end{subtable}
\hfill
\begin{subtable}[t]{0.24\linewidth}
\centering
\caption{C3: Pump.fun-Centric Trading}
\begin{tabular}{>{\raggedright\arraybackslash}p{0.55\linewidth} >{\raggedleft\arraybackslash}p{0.4\linewidth}}
\toprule
\textbf{Token Mint} & \textbf{Txs \# | \%} \\
\midrule
WSOL & 22,894 | 10.93\% \\
SamAI (Pump.fun) & 16,456 | 7.85\% \\ 
Breastcoin (Pump.fun) & 8,234 | 3.93\% \\ 
Cameo (Pump.fun)& 4,790 | 2.29\% \\ 
USDC & 3,699 | 1.77\% \\
PIMP (Pump.fun)& 2,160 | 1.03\% \\ 
USD1 & 1,751 | 0.84\% \\
Bella (Pump.fun)& 1,206 | 0.58\% \\ 
PAYAI (Pump.fun)& 1,198 | 0.57\% \\ 
PFP (Pump.fun)& 1,145 | 0.55\% \\ 
\bottomrule
\end{tabular}
\end{subtable}
\vspace{3pt}
\begin{minipage}{\textwidth}
\footnotesize
\textsuperscript{a} {The token is named \textit{1 coin can change your life}, with \textit{1} used as its abbreviation. 
}
\end{minipage}
\end{table*}

\begin{table}[ht]
\centering
\footnotesize
\caption{Top transfer recipients across clusters after merging by public name or address prefix for unlabeled accounts.}
\label{tab:top_recipients_by_cluster}
\resizebox{0.7\linewidth}{!}{%
\begin{tabular}{lp{2.6cm}rc}
\toprule
 & \textbf{Recipient (Merged)} & \textbf{Total Transfer | \%} & \textbf{\# Bot Addresses} \\
\midrule
\multirow{6}{*}{\parbox{3.5cm}{\textbf{C0: Multi-Venue MEV}}}
& Pump.fun Protocol Fee & 26,682,284 | 0.56\% & 2 \\
& Jitotip& 17,720,468 | 0.37\% & 4 \\
& Others &  &  \\
\cmidrule(lr){2-4}
& \hspace{0.5em}{ $\triangleright$ ReNT...j4aY} & 4,448,417,434 | 93.25\% & 4 \\
& \hspace{0.5em}{ $\triangleright$ FEnA...596o} & 181,568,661 | 3.81\% & 1 \\
& \hspace{0.5em}{ $\triangleright$ 7tDT...pRcZ} & 50,000,000 | 1.05\% & 1 \\
& \hspace{0.5em}{ $\triangleright$ FAST...wN7M} & 28,045,984 | 0.59\% & 3 \\
\midrule
\multirow{5}{*}{\parbox{3.5cm}{\textbf{C1: Venue-Specialized MEV}}}
& Jitotip & 34,965,963,665 | 62.65\% & 10 \\
& Others &  &  \\
\cmidrule(lr){2-4}
& \hspace{0.5em}{ $\triangleright$ astra} & 12,423,059,423 | 22.26\% & 4 \\
& \hspace{0.5em}{ $\triangleright$ flash} & 1,225,770,404 | 2.20\% & 9 \\
& \hspace{0.5em}{ $\triangleright$ 7tDT...pRcZ} & 500,000,000 | 0.90\% & 1 \\
\midrule
\multirow{6}{*}{\parbox{3.5cm}{\textbf{C2: Aggregator-Centric MEV}}}
& Jitotip & 133,150,141,077 | 52.88\% & 11 \\
& solanamevbot & 4,700,495,258 | 1.87\% & 1 \\
& Helius & 44,176,347 | 0.02\% & 3 \\
& Others &  & \\
\cmidrule(lr){2-4}
& \hspace{0.5em}{ $\triangleright$ BFP8...WGEu} & 103,640,000,000 | 41.16\% & 1 \\
& \hspace{0.5em}{ $\triangleright$ astra} & 10,147,270,614 | 4.03\% & 6 \\
\midrule
\multirow{6}{*}{\parbox{3.5cm}{\textbf{C3: Pump.fun-Centric Trading}}}
& Trojan & 1,752,718,250,257 | 61.26\% & 96 \\
& Jitotip & 290,865,285,692 | 10.17\% & 100 \\
& Pump.fun & 185,504,355,820 | 6.48\% & 88 \\
& Others &  &  \\
\cmidrule(lr){2-4}
& \hspace{0.5em}{ $\triangleright$ FbWy...V4wi} & 120,325,058,212 | 4.21\% & 3 \\
& \hspace{0.5em}{ $\triangleright$ 6ggZ...zCFV} & 106,841,881,088 | 3.73\% & 2 \\
\bottomrule
\end{tabular}
}
\end{table}

\begin{table}[t]
\centering
\caption{\hl{Human validation for the dominant category assignment of Solana bot repositories across the 15 categories.}}
\label{tab:repo_validation}
\resizebox{0.6\linewidth}{!}{%
\begin{tabular}{lccc}
\toprule
\textbf{Category} & \textbf{Sampled ($N$)} & \textbf{Confirmed ($n$)} & \textbf{Accuracy (\%)} \\
\midrule
Order Execution and Management        & 14 & 13 & 92.9 \\
Sniping                                & 13 & 11 & 84.6 \\
Arbitrage                              & 8  & 7  & 87.5 \\
Transaction Ordering Exploitation      & 8  & 7  & 87.5 \\
Messaging Integration and Alerts       & 7  & 7  & 100.0 \\
Copy Trading                           & 7  & 7  & 100.0 \\
DEX Swap Routing and Integration       & 7  & 7  & 100.0 \\
Monitoring and Alerting                & 6  & 5  & 83.3 \\
Wash Trading                           & 6  & 6  & 100.0 \\
Portfolio and Risk Management          & 5  & 4  & 80.0 \\
NFT Minting and Marketplace Trading    & 5  & 5  & 100.0 \\
Liquidity Provision and Yield Farming  & 4  & 3  & 75.0 \\
On-chain Data Pipeline                 & 4  & 3  & 75.0 \\
Token Issuance and Administration      & 3  & 3  & 100.0 \\
Wallet Management                      & 3  & 3  & 100.0 \\
\midrule
\textbf{Overall} & \textbf{100} & \textbf{91} & \textbf{91.0} \\
\bottomrule
\end{tabular}
}
\end{table}

\clearpage
\twocolumn

\bibliographystyle{IEEEtran}
\bibliography{solana_bot}

\end{document}